\documentclass[12pt,preprint]{aastex}

\slugcomment{Accepted to appear in \apj; RAP \#475}

\shorttitle{NGC 1365.}
\shortauthors{Z\'anmar S\'anchez \etal}

\newcommand\ha{\hbox{H$\alpha$}}
\newcommand\chisq{\hbox{$\chi^2$}}
\newcommand\kpc{\hbox{$\rm{kpc}$}}
\newcommand\kms{\hbox{$\rm{km}~\rm{s}^{-1}$}}

\def\eg{{\it e.g.}}
\def\etal{{\it et al.}}

\def\ie{{\it i.e.}}

\long\def\Ignore#1{\relax}

\begin{document}

\title{Modeling the Gas Flow in the Bar of NGC 1365}

\author{R. Z\'anmar S\'anchez}

\author{J. A. Sellwood}
\affil{Department of Physics and Astronomy, Rutgers University, \\
    136 Frelinghuysen Road, Piscataway, NJ 08854}
\email{zanmar,sellwood@physics.rutgers.edu}

\author{B. J. Weiner}
\affil{Steward Observatory, Department of Astronomy, University of Arizona, \\
     933 N. Cherry Ave., Tucson, AZ 85721}
\email{bjw@as.arizona.edu}

\and

\author{T. B. Williams}
\affil{Department of Physics and Astronomy, Rutgers University, \\
    136 Frelinghuysen Road, Piscataway, NJ 08854}
\email{williams@physics.rutgers.edu}

\begin{abstract}
We present new observations of the strongly-barred galaxy NGC~1365,
including new photometric images and Fabry-Perot spectroscopy, as well
as a detailed re-analysis of the neutral hydrogen observations from
the VLA archive. We find the galaxy to be at once remarkably
bi-symmetric in its I-band light distribution and strongly asymmetric
in the distribution of dust and in the kinematics of the gas in the
bar region. The velocity field mapped in the \ha\ line reveals bright
HII regions with velocities that differ by 60 to $80\;$\kms\ from
that of the surrounding gas, which may be due to remnants of infalling
material.  We have attempted hydrodynamic simulations of the bar flow
to estimate the separate disk and halo masses, using two different
dark matter halo models and covering a wide range of mass-to-light
ratios ($\Upsilon$) and bar pattern speeds ($\Omega_p$).  None of our
models provides a compelling fit to the data, but they seem most
nearly consistent with a fast bar, corotation at $\sim1.2r_B$, and
$\Upsilon_ I \simeq 2.0 \pm 1.0$, implying a massive, but not fully
maximal, disk.  The fitted dark halos are unusually concentrated, a
requirement driven by the declining outer rotation curve.
\end{abstract} 


\section{Introduction}
The centrifugal balance of the circular flow pattern in a
near-axisymmetric spiral galaxy yields a direct estimate of the
central gravitational attraction as a function of radius.  However,
the division of the mass giving rise to that central attraction into
separate dark and luminous parts continues to prove challenging.  The
radial variation of the circular speed simply does not contain enough
information to allow a unique decomposition between the baryonic mass,
which has an uncertain mass-to-light ratio, $\Upsilon$, and the dark
halo, whose density profile is generally described by some adopted
parametric function \citep{albada3198, LF89, BSK04}.

Predictions for $\Upsilon$ from stellar population synthesis models
that match broad-band colors \citep[e.g.][]{Bell03} are useful, but
not precise.  Despite intense effort, they are still sufficiently
uncertain to be consistent with both maximum and half-maximum disk,
which is the range of disagreement \citep[e.g.][]{Sackett97,
Bottema97, S99Rutgers}.  \citet{McGaugh05} argues that the values can
be refined by minimizing the scatter in the Tully-Fisher and/or mass
discrepancy-acceleration relation.

A number of dynamical methods have been employed to break the
disk-halo degeneracy.  \citet{Casertano83}, \citet{bosma98}, and
others have suggested that the slight decrease in orbital speed near
the edge of the optical disk of a bright galaxy -- the ``truncation
signature'' -- could be used as an indicator of disk $\Upsilon$, but
in practice it does not provide a tight constraint.  \citet{ABP87} and
\citet{Fuchs03} attempt to constrain the disk mass using spiral
structure theory.  \citet{Bottema97} and \citet{verheijen04} measure
the vertical velocity dispersion of disk stars in a near face-on
galaxy, which they assume has the same mean thickness of similar
galaxies seen edge-on \citep{KvdKdG}, to constrain the disk mass.  A
similar approach is reported by \citet{CD04} using velocity
measurements of individual planetary nebulae.

One of the most powerful, although laborious, methods for barred
galaxies was pioneered by Weiner, Sellwood \& Williams (2001), who
made use of the additional information in the driven non-circular
motions caused by the bar.  By modeling the observed non-axisymmetric
flow pattern of the gas in a 2-D velocity map, they were able to
determine the mass-to-light ratio of the visible disk material.  They
found that the luminous disk and bar contributed almost all the
central attraction in NGC~4123 inside $\sim 10\;$kpc, requiring the
dark halo to have a very low central density.  \citet{Ben3095} reports
a similar result for a second case, NGC 3095.  The method has also
been applied by \citet{perez04} for several barred galaxies and by
\citet{kranz03} who modeled motions caused by spiral arms.
\citet{BEG03} present a similar study for the Milky Way.  Earlier
studies \citep[\eg][]{DA83} did not attempt to separate the disk from
the dark matter halo \citep[see][for a review]{SW93}.  Here, we apply
the \citep{wein2} method to the more luminous barred galaxy NGC~1365
in the Fornax cluster.

As one of the most apparently regular, nearby barred spiral galaxies
in the Southern sky, NGC~1365 was selected by the Stockholm group for
an in-depth study \citep[see \eg][]{Lindblad1365}.  Hydrodynamic
models of the bar flow pattern were already presented by
\citep{Linlinatha}, based mainly on the velocities of emission-line
measurements from many separate long-slit observations.  J\"ors\"ater
\& van Moorsel (1995, hereafter JvM95) present a kinematic study using
the 21 cm line, which suggests that the galaxy is somewhat asymmetric
in the outer parts, where the shape of the rotation curve is hard to
determine.
\citet{sandqvist} find substantial amounts of molecular gas, but only
within 2~kpc of the nucleus, which is resolved in interferometric
observations \citep{Sakamoto07} into a molecular ring in
the center plus a number of CO hot spots.
\citet{Galliano05} have found previously unknown MIR sources in the
inner 10\arcsec\ around the AGN. They are able to correlate some of
these MIR sources with radio sources, which they interpret in terms of
embedded star clusters because of the lack of strong optical
counterparts.
\citet{JCA97} present H-band photometry of the bright inner disk,
finding an elongated component in the central region suggesting that
the NGC~1365 is a double-barred galaxy, although they note that the
light in this component is not as smooth as in their other nuclear bar
cases. \citet{LSKP} also classify it as a double barred galaxy.
However, \citet{Emsellem01} and \citet{Erwin04} argue against a
nuclear bar, citing an HST NICMOS image which resolves the feature
into a nuclear spiral.
\citet{Emsellem01} also present stellar kinematics from slit spectra
using the $^{12}$CO bandhead.  They propose a model for the inner 2.5
kpc of NGC 1365 consisting of a decoupled nuclear disk surrounded by
spiral arms within the inner Lindblad resonance (ILR) of the primary bar.
\citet{Beck05} observed NGC 1365 in radio continuum at
9\arcsec-25\arcsec\ resolution and find radio ridges roughly
overlapping with the dust lanes in the bar region. They propose that
magnetic forces can control the flow of gas at kiloparsec scales.

Here we present new photometric images, a full 2-D velocity map of the
\ha\ emission, and a reanalysis of the neutral hydrogen data from
JvM95.  We also compare many hydrodynamic models to these new data in
an effort to determine the separate disk and dark halo masses in this
galaxy.

\section{Observations}
The \citet{wein2} method requires both broad band photometry, to
estimate the distribution of visible matter, and a high spatial
resolution velocity map to determine the projected non-axisymmetric
flow pattern in the barred region.  While not absolutely essential,
knowledge of the rotation curve at larger radii is helpful to estimate
the total central attraction.  Velocity maps using the 21cm line of
neutral hydrogen have lower spatial resolution but extend to larger
radii, and are therefore an ideal complement to the optical data.

Here we adopt the distance to NGC 1365 of 18.6 Mpc, as deduced from
Cepheid variables by \citet{Madore}.  At this distance, $1\arcsec$
corresponds to $90.2\;$pc.

\begin{figure}[p]
\centering
\includegraphics[angle=270,scale=0.55,viewport=0 0 510 440,clip]{Iband.ps}\\

\includegraphics[angle=270,scale=0.32,viewport=0 0 550 730,clip]{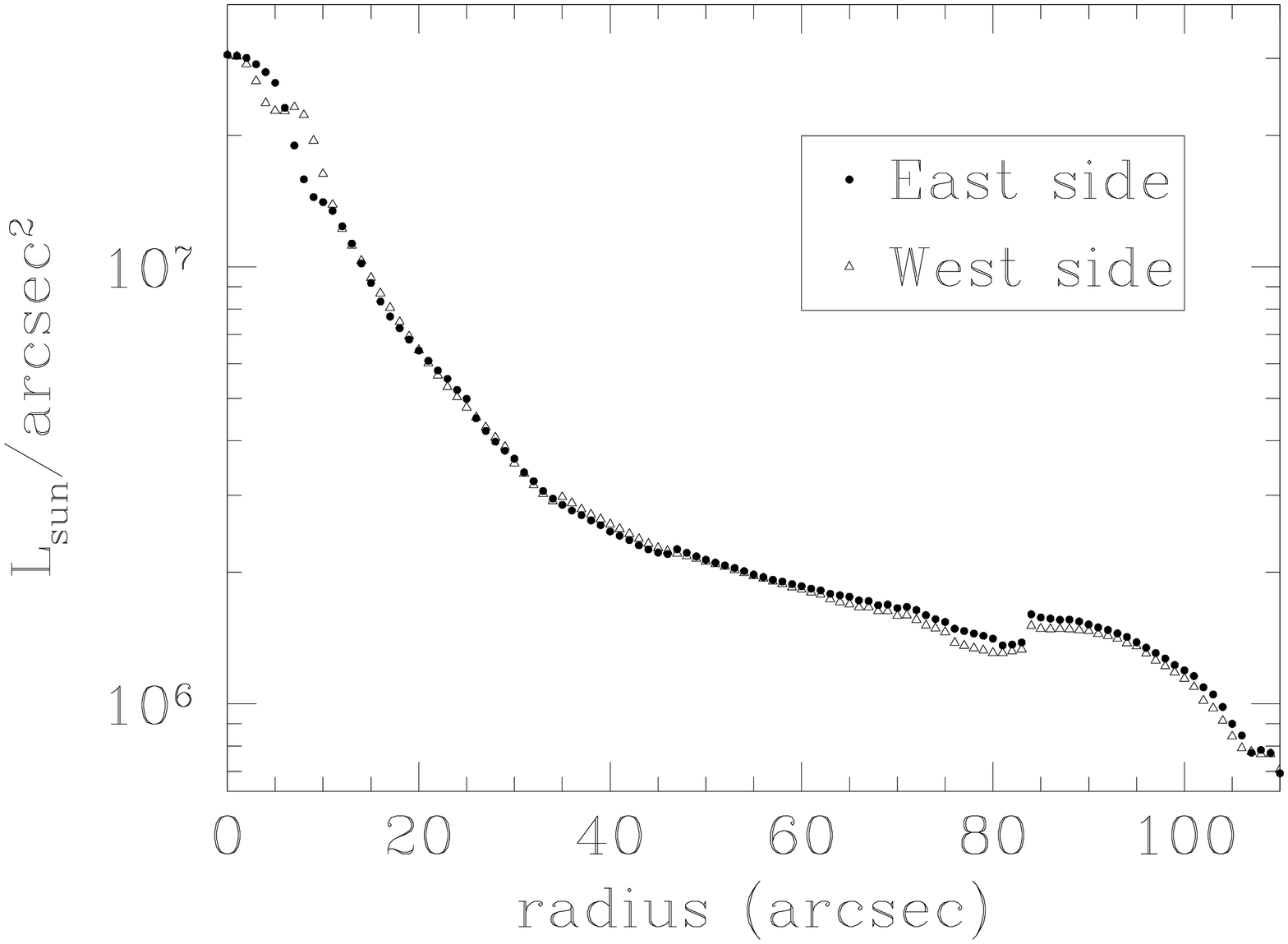}\\
\caption{\footnotesize \textit{Upper:} A $700\arcsec$ square region
with the I-Band image of NGC 1365 showing intensity on a log scale.
The intensity range does not represent the full range and has been
chosen to reveal the spiral and bar structures most clearly (\ie\ the
center of the galaxy has a luminosity of $\sim 3\times\ 10^{7}
L_{\sun}/arcsec^2$). A white contour with intensity $1.2\times\ 10^{6}
L_{\sun}/arcsec^2 $ has been plotted to outline the bar. The black
contour is the same as the white but rotated by $180\degr$; the
similarity of the two contours shows the remarkable 2-fold symmetry of
the bar.  Subsequent figures show this isophote for reference drawn
from a 2-fold rotationally averaged image.  \textit{Lower:} Surface
brightness profiles for the east and west sides of the bar estimated
independently. The length of the semimajor axis of the bar is
100\arcsec.}
\label{ibandfig}
\end{figure}

\subsection{Surface Photometry}
Observations of NGC 1365 were made on the night of 1999 Jan 24 with
the Swope 1~m telescope at Las Campanas Observatory.  The LCO Tek\#5
$2048\times2048$ pixel CCD was used with a pixel scale of
$0.{\arcsec}70$ and a field of view of $24{\arcmin}$.  We obtained
$3\times10$ minute offset exposures in each of the V \& I filters.
The seeing was approximately $1.{\arcsec}6$.  We reduced the images
with IRAF\footnote{IRAF is distributed by NOAO, which is operated by
AURA, Inc., under a cooperative agreement with the National Science
Foundation.} with the standard procedure: subtraction of overscan,
bias, and flat-fielding with twilight flats.  The LCO CCD detector
exhibits fringing and illumination gradients in the I filter.  These
were removed by constructing a supersky image by combining all the
available I-band images taken throughout the night and subtracted from
NGC 1365 I-band image.  Photometric standard stars from \citet{Lan92}
were observed at several times during the night and used to derive
extinction coefficients.

The resulting calibrated I-band image is presented in the upper panel
of Figure~\ref{ibandfig}.  We have drawn a single isophotal contour
and redrawn the same contour rotated through 180$^\circ$ in order to
show the symmetry of the bar light distribution.  This particular
isophote shows the approximate size of the bar, as determined below.
We see that the east side of the bar is marginally fatter, at this
isophote level, than is the west, but the shape of the bar is
remarkably symmetric.  In subsequent figures we draw, as a reference,
a slightly smoothed version of this isophote from the 2-fold
rotationally averaged I-band light distribution.

\begin{figure}[t]
\centering
\includegraphics[angle=270,scale=0.7]{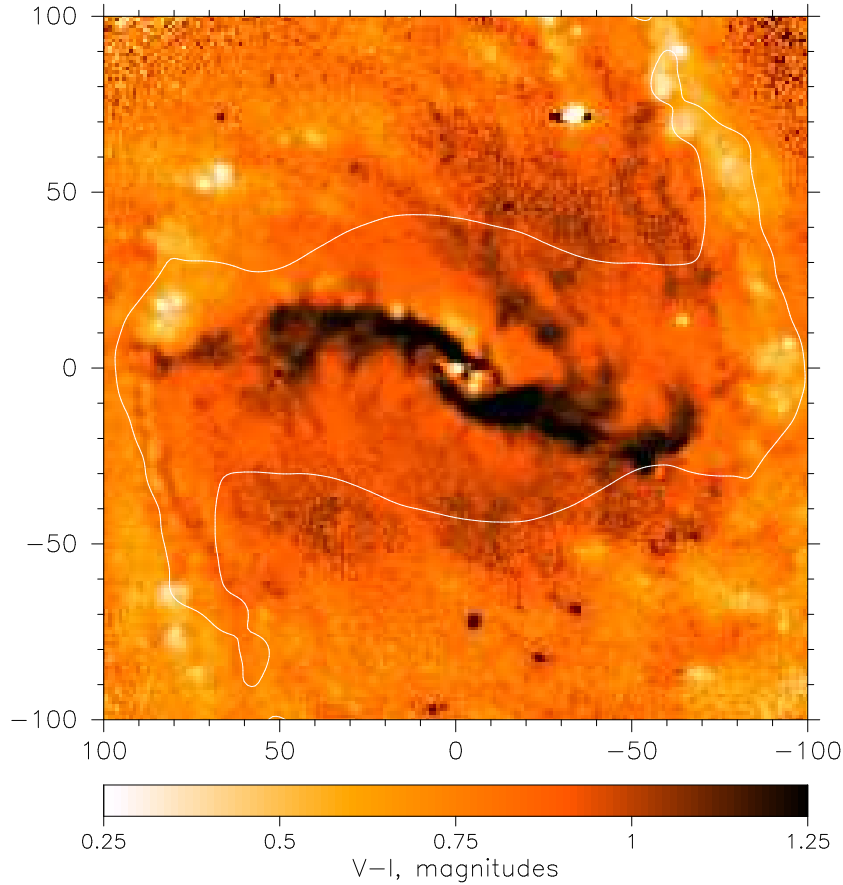}
\caption{\footnotesize A V-I color map of the central 200\arcsec\
square region of NGC 1365.  Dust lanes are clearly visible on the
leading sides of the bar.  The isophote from Fig.~\ref{ibandfig} is
included for reference.  Note the absence of symmetry in the dust
lanes; that on the west side lies farther from the bar major-axis.}
\label{dustfig}
\end{figure}

The lower panel in Figure \ref{ibandfig} presents independent
estimates of the surface brightness profile for the left and right
side of the bar, constructed as follows.  We ran the IRAF-ellipse tool
on a version of the I-band image smoothed to 3\arcsec, to find a set
of isophotal ellipses with position angle and ellipticity as free
parameters, but with a fixed center.  We then used the same generated
ellipses on the original I-band image to find the mean intensity for
the east and west sides independently.  This analysis reveals that the
light distribution is indeed highly symmetric.  The apparent
discontinuity in the light profiles at $r\sim85\arcsec$ is due to
variations in the fitted ellipticity caused by a number of bright HII
regions near the end of bar.

The strong dust lanes in the bar, and other places, are easily visible
in the V-I color map of NGC 1365 (Figure~\ref{dustfig}).  As usual,
the dust lanes are on the leading side of the bar (which rotates
clockwise if the spiral arms are assumed to trail).  Despite the
evident 2-fold symmetry of the bar, the dust lanes are clearly not
symmetric; the strongest dust features on the west side lie farther
towards the leading edge of the bar than those on the east.  This
asymmetric extinction could be the cause of the mild asymmetry in the
bar light (Fig.~\ref{ibandfig}) noted above.

Since the NE side of the galaxy is approaching while the SW receding,
trailing spiral arms imply that the NW of the galaxy is tipped towards
us.  Assuming the redder regions in the color map indicate diffuse
dust, there appears to be more extinction on the near (NW) side than
on the far (SE) side, which is consistent with expectations for a
moderately inclined disk \citep[\eg][\S4.4.1]{BM98}. \cite{Holwerda05}
present a detailed study of dust extinction in part of NGC~1365.

We deproject the galaxy to obtain a face-on surface brightness
profile, adopting the projection geometry indicated by the kinematic
maps presented in \S\ref{rotcursec}: position angle (hereafter PA)
$=220^{\circ}$ and inclination $i = 41^\circ$.  We estimate
$R_{23.5}=348\arcsec \simeq 31.4\;$\kpc\ in the I-band, which is
comparable to B magnitude $R_{25}=337\arcsec$ \citet{RC3}, and an
exponential scale length for the disk of $r_d \simeq 6.5\;$\kpc.  We
estimate the total magnitude within $R_{23.5}$ to be 8.2 in I and 9.4
in V, without any extinction corrections.  For our modeling of the
potential from the I band, we apply an internal extinction correction
$A_{\rm int}=-1.0\log(b/a)$ from \citet{giovanelli94} and a galactic
extinction from \citet{schlegel} for a total correction of $-0.16$ at
I.

We estimated the size of the bar by fitting ellipses to the
deprojected I-band image at a number of isophote levels.  Ideally, one
hopes for an abrupt change in the ellipticity and PA of the radial
profiles at the transition between the bar and the disk.  The spiral
arms, however, force the PA to vary continuously and broaden the
ellipticity peak somewhat \citep{wozniak95}.  We estimate the
deprojected bar semi-major axis to be $114\arcsec \la r_B \la
127\arcsec$, where the lower limit is the radius of maximum
ellipticity, 0.63, and the upper limit is radius at which the PA
begins to change sharply.  This range is in agreement with the value
of $r_B = 120\arcsec \pm 10\arcsec$ estimated by \cite{Linlinatha}
using Fourier moment decomposition, and hereafter we adopt their value
$r_B=120\arcsec=10.8\;\kpc$ as the bar semi-major axis.  While
physically large, $r_B \simeq 1.66r_d$, which is typical
\citep{Erwin05}.

In this work, we use our I-band image as a measure of the light from
the old disk stars.  Although far less problematic than in the V-band,
some residual dust obscuration slightly attenuates the I-band light.
As the opacity of dust is still lower in infrared light, it might be
argued that a better estimate of the underlying stellar light
distribution could be obtained from J, H, or K-band images.  However,
the NIR bands are not a panacea as the light in these bands is more
seriously affected by AGB stars and hot dust in HII regions, as may be
seen by the lumpiness of the 2MASS \citep{TMASS} K-band image.  In
addition, no available NIR detector is large enough to cover this
galaxy well out to the sky without mosaicking, and no existing near
infrared image is deep enough (\eg\ 2MASS) to yield the accurate
measure of the outer disk surface brightness profile we require.
Furthermore, \citet{dej96} shows that the dependence of the I-band
surface brightness on age and star formation history is only slightly
stronger than for the K-band.  We therefore adopt our I-band image as
the best available estimator of the underlying disk light down to very
low surface brightness levels.

\subsection{Fabry-Perot Imaging Spectroscopy}
\label{fabrysec}
NGC 1365 was observed in the \ha\ line on the nights of 1993 November
3-4 with the CTIO 1.5m telescope using the Rutgers Imaging Fabry-Perot
interferometer.\footnote{CTIO is operated by AURA under contract to
the NSF.}   A Tektronix $1024\times1024$ CCD detector was used with
$0.{\arcsec}98$ pixels.  The field of view of the etalon was
$7.{\arcmin}8$ in diameter.  Hourly calibrations were taken in order
to correct for temporal drifting of the wavelength zero point and
optical axis center.

The first night of the observations was photometric but the second
night was plagued with intermittent cloud, and three exposures had to
be discarded.  The remaining NGC 1365 observations consist of 16
separate 10-minute exposures spanning 17.4 \AA\ in steps of $\sim$1.2
\AA\ (54 \kms) and covering a range of velocities from 1010 to 2031
km/s.  The reduction methods are similar to those described elsewhere
\citep{palunas00,wein1}.  The images were reduced with IRAF: overscan
subtraction, bias, and flat fielding.  After the removal of cosmic
rays, the images were spatially registered and sky subtracted to build
the data cube.  The typical seeing as measured from foreground stars
was around $2.4\arcsec$.  Sky transparency variations occurred
throughout the observation and are a major source of uncertainties in
line profiles \citep{ted84}.  We measure the flux of brightest,
isolated foreground star in the field of view and scale the frames to
a common transparency.  Corrections never exceed 3\%.

Four frames of our cube seem to be mildly contaminated by sky lines
$\lambda=6604.13$ and $\lambda=6596.64$ \AA\ produced by air OH
rotation-vibration transitions \citep{sky}.  A model of the sky ring
can be easily constructed because the spectral width of sky lines is
very narrow and we should recover only the unresolved instrumental
Voigt profile.  Nevertheless, the peak intensity of the sky ring is
hard to determine because of the confusion with the light of the
embedded galaxy.  We tried a range of peak intensity values and
convinced ourselves that it is no more that 4-5 counts above the
background level. Subtraction of these sky rings eliminated spurious
fits in the faint outskirts of the galaxy and had no noticeable effect
on the velocity fits within the galaxy.

The Fabry-Perot has a spectral resolution of 2.5 \AA\ at \ha\ or
$\rm{FWHM} \simeq 150\;$\kms.  The spectral profile was approximated
with a Voigt function with Gaussian $\sigma_{G}= 21.6\;$\kms\ and
Lorentzian $\sigma_{L}= 61.4\;$\kms.  In order to improve the S/N of
our line profiles, we combined data from adjacent pixels using a
varying Gaussian kernel of up to $5 \times 5$ pixels for which the
FWHM is adapted depending on the strength of the line.  We use only
the central pixel when the line is very strong, but bin pixels $3
\times 3$ (FWHM=1.6\arcsec) for lines of intermediate strength, and
use a $5 \times 5$ box (FWHM=3.3\arcsec) for very weak emission.

\begin{figure}[t]
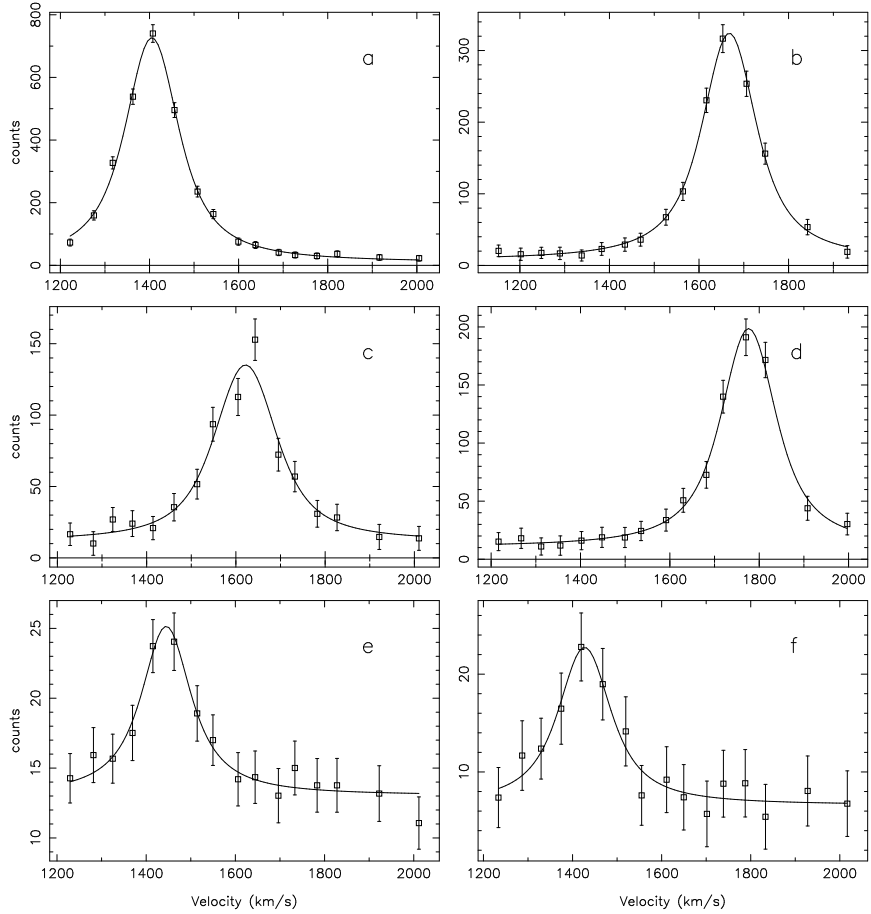

\centering
\includegraphics[angle=270,scale=0.25,viewport=0 0 440 650,clip]{prof_a.ps}
\includegraphics[angle=270,scale=0.25,viewport=0 20 440 650,clip]{prof_b.ps}\\

\includegraphics[angle=270,scale=0.25,viewport=0 0 440 650,clip]{prof_c.ps}
\includegraphics[angle=270,scale=0.25,viewport=0 20 440 650,clip]{prof_d.ps}\\

\includegraphics[angle=270,scale=0.25,viewport=0 0 480 650,clip]{prof_e.ps}
\includegraphics[angle=270,scale=0.25,viewport=0 20 480 650,clip]{prof_f.ps}\\

\caption{\footnotesize Example line profiles.  \textit{a.)} and
\textit{b.)}  Bright HII region on spiral arm.  \textit{c.)} and
\textit{d.)} Bright HII regions from east and west side of the bar
respectively; \textit{c.)}  is from the region labeled R2 in
Fig.~\ref{velfig}.  \textit{e.)} and \textit{f.)} Diffuse emission
from east side of the bar.}
\label{profilesfig}
\end{figure}

Fabry-Perot images are not strictly monochromatic because of the
angular dispersion of the instrument.  For every pixel, the wavelength
is calculated from a quadratic variation with the radial distance from
the optical axis determined from calibration lamp exposures.  With the
fluxes and wavelengths for every pixel, a Voigt function of five
parameters can be fitted with least-squares techniques.  After some
experimentation, we decided to fix the Lorentzian width to the
instrumental value and fitted the four remaining parameters: center
wavelength (velocity), Gaussian line width, peak intensity of the
emission and the continuum level.  Figure~\ref{profilesfig} shows a
selection of resultant emission line profiles: (a) and (b) are of very
bright HII regions in the spiral arms, (c) and (d) are of bright HII
regions in the eastern and western side of the bar respectively, while
(e) and (f) are of diffuse emission from the east side of the bar.

\begin{figure}[t]
\centering
\includegraphics[angle=270,scale=0.4]{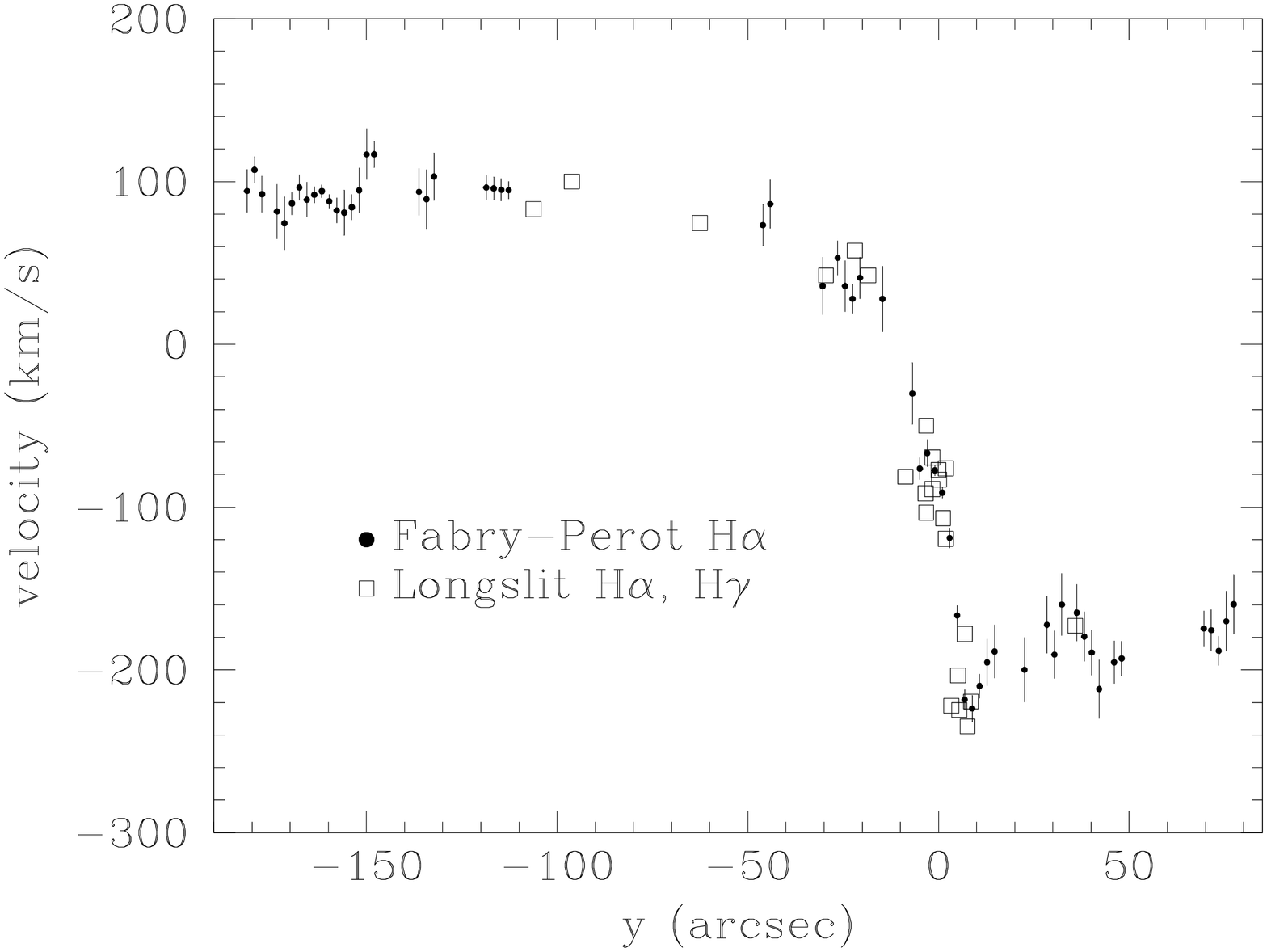}\\
\caption{\footnotesize Comparison between line-of-sight velocities
from Fabry-Perot data (filled points) and the long slit measurements
(open symbols) from \cite{Lindslit96}.  Data from their Fig.~2e are
based on the \ha\ and H$\gamma$ lines from a slit positioned in the
N-S direction and centered on the bright HII region L33.}
\label{compare}
\end{figure}

\begin{figure}[t]
\centering
\includegraphics[angle=270,scale=0.8]{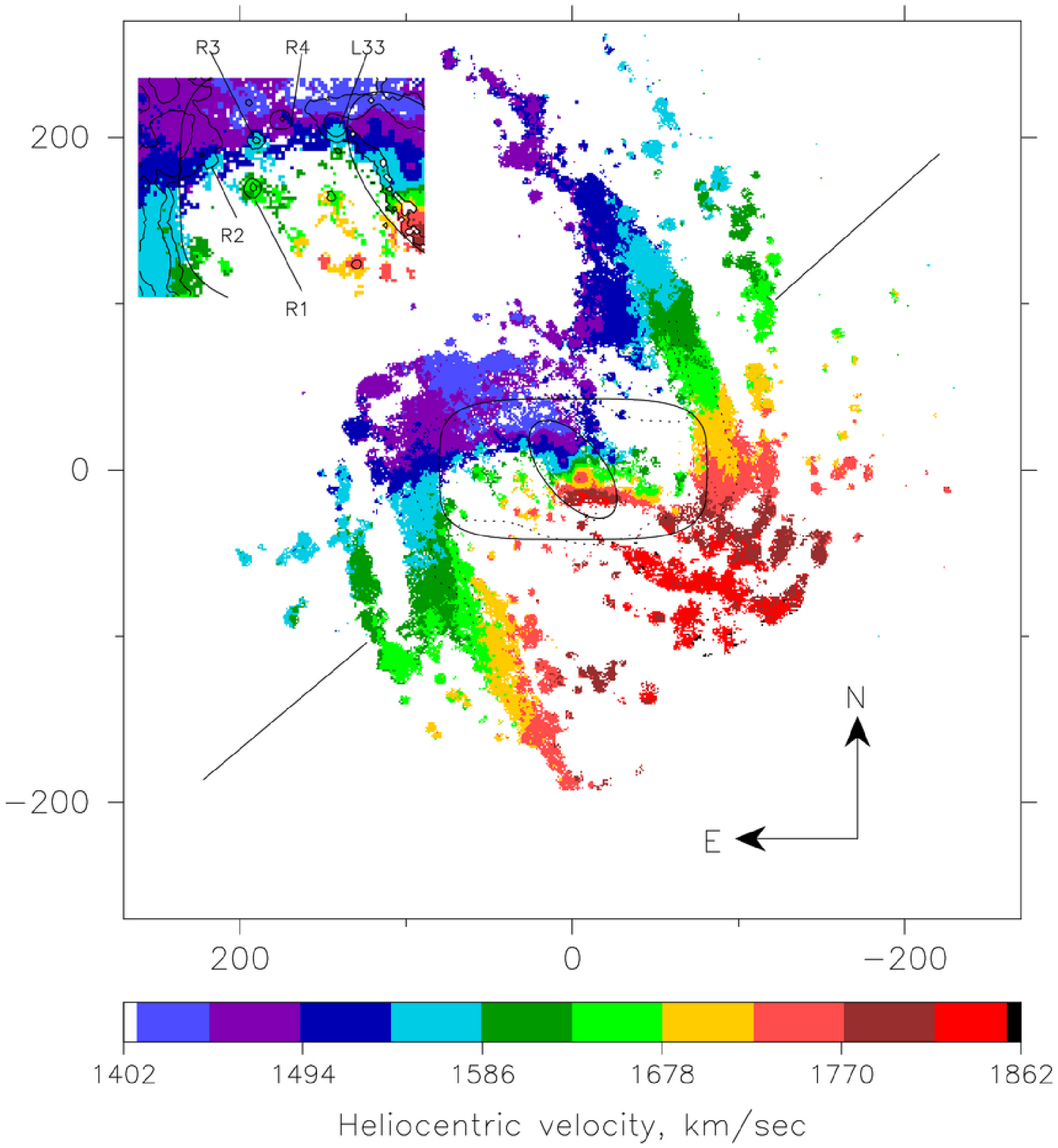}
\caption{\footnotesize The velocity map of NGC 1365 from Fabry-Perot
observations of the \ha\ emission line.  The color map has been binned
in 10 colors to outline velocity contours.  The dotted line shows the
contour from Fig.~\ref{ibandfig}.  The closed solid curves are
described in \S\ref{comparsect}.  The upper-left inset shows an
enlargement of the east side of the bar with contours of intensity to
show the positions of several labeled bright HII regions.}
\label{velfig}
\end{figure}

Fits to every pixel yield maps of velocity, line strength, line width,
continuum level and the estimated uncertainty in each quantity.
Figure \ref{velfig} presents the velocity map; regions where the
velocity uncertainty exceeds $20\;$\kms\ are left blank.  We
illustrate the velocities using 10 colors only so that the interface
between two colors marks an isovelocity contour.  The dotted outline
is the reference isophote of the bar while the two straight solid
lines in the outer regions lie along the estimated (see \S
\ref{rotcursec}) minor axis of the galaxy and help to identify kinks
in the velocity field inside the bar.

Figure \ref{compare} presents a comparison between our Fabry-Perot
velocity measurements and the long slit measurements using \ha\ and
H$\gamma$ from \cite{Lindslit96}, taken from their Fig 2.e.  Their
slit was centered on bright HII region L33 and runs along the
South-North direction with the zero point of the velocity set to 1630
km/s.  We also find very good agreement (within $1-\sigma$) between
our measured velocities for other bright HII regions L2, L3, L4, L29,
L32, as labeled by \citet{alloin}, and HII region containing
supernova 1983V.  Our estimates for the nucleus and L1 differ from
theirs by $\sim40\;$km/s, which is not surprising since NGC 1365 is a
Seyfert galaxy and several authors
\citep{Phillips83,edmunds88,Lindslit96} have detected splitting of the
high excitation line [OIII].  Double-valued velocities are thought to
arise from a disk component and a bipolar hollow conical outflow, as
first proposed by \citet{Phillips83}.

The velocity map reveals steep gradients in the bar region, especially
on the leading edges, where the dust lanes are also found, as has been
reported previously \citep[][and references
therein]{jorsater84,Lindslit96,teuben86}.  These steep gradients are
strongly asymmetric as is clear in our velocity map.  The velocity
jumps on the west side of the bar extend farther to the leading side
of the bar, consistent with the asymmetric arrangement of the dust
lanes.  The asymmetry can also be seen in Fig.~4 of
\citet{Lindslit96}, although based on a rather sparse spatial coverage
(interpolated map from 35 slits).

The line profile (Fig.~\ref{profilesfig}c) of the bright HII region on
the east side of the bar labeled R2 in Fig.~\ref{velfig} (offset from
the nucleus $x=\Delta\alpha \cos \delta=54.60\arcsec$,
$y=\Delta\delta=13.88\arcsec$), is not well fitted by a single
broadened velocity, but seems to be more complex.  Furthermore, the
mean fitted velocity at this location differs substantially from that
of the faint emission in the surrounding pixels (see inset of
Fig.~\ref{velfig}).  Other bright HII regions on the east side of the
bar (\eg\ R1 at $x=45.90\arcsec$, $y=20.64\arcsec$ and R3 at
$x=55.33\arcsec$ $y=-0.84\arcsec$) show similar anomalies.

In order to quantify the velocity difference between these three
bright regions and that of the surrounding diffuse gas, we fit a
bi-linear velocity gradient to the diffuse emission over a small
square patch ($20\times20$ pixels) surrounding each bright region.
The interpolated velocity at the center of the HII regions differs by
between 60 \& 80~km/s from the measured value.  These complicated
profiles may result from disturbance by infalling gas, as discussed in
\S\ref{asymmsec}.

\subsection{HI radio observations}
NGC 1365 was observed at the VLA in the HI 21 cm line in 1986 (see
JvM95) for 48 hours in three different configurations: BnA, CnB
and DnC.  The observations were collected on 11 different days;
further details of the observations are given in JvM95.  Motivated by
improvements in numerical routines in AIPS and by the puzzling
declining rotation curve reported by these authors, RZS spent a
summer at NRAO\footnote{The NRAO is operated by Associated
Universities, Inc., under a cooperative agreement with National
Science Foundation.} re-analyzing these observations of NGC 1365
under the supervision of Gustaaf van Moorsel.

Data from the different days were reduced independently and then
combined in the UV plane using standard AIPS calibration procedures
(DBCON, UVLIN).  The resulting combined beamsize was $10.3\arcsec
\times 9.7\arcsec$.  Only 31 channels with a corresponding velocity
resolution of $20.84\;$\kms\ were used.

\begin{figure}[t]
\centering
\includegraphics[angle=270,scale=0.7]{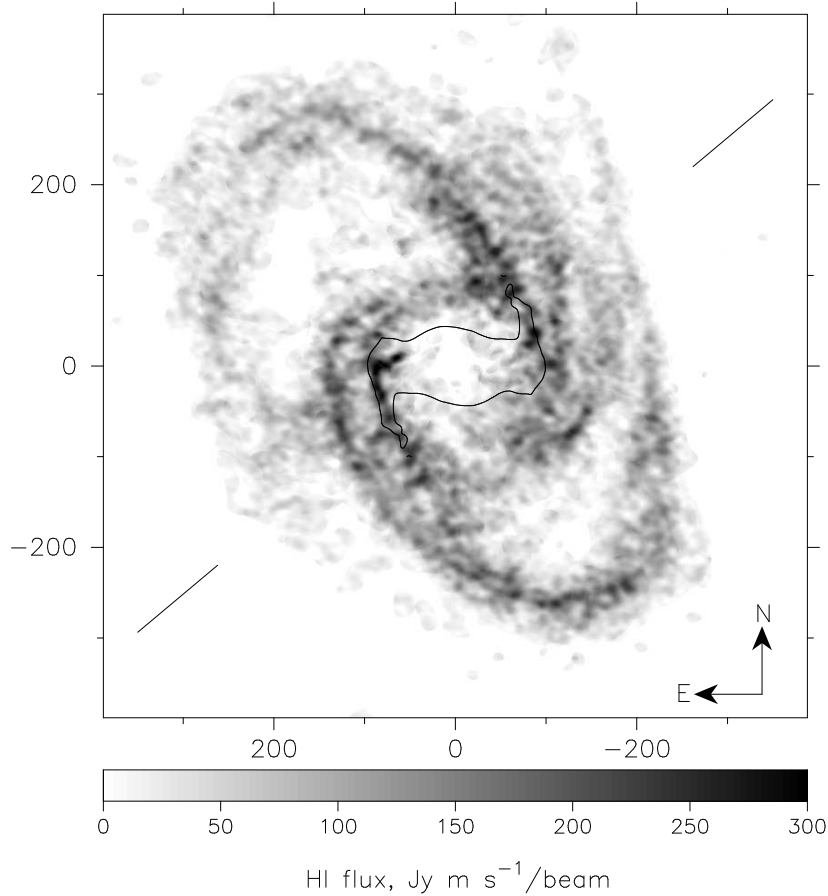}
\caption{\footnotesize The distribution of neutral hydrogen, from the
velocity-integrated 21-cm surface brightness. The beam size is
$10.3\arcsec \times 9.7\arcsec$}
\label{HIint}
\end{figure}

An image cube was created with the task IMAGR using robust weighting
\citep[see][]{brigs95}.  Natural and uniform weighting are controlled
in AIPS by the robustness parameter and we explored the range from
$-4$ (uniform weighting) to 4 (natural weighting) by measuring the RMS
of an empty box in the resulting map and by checking the dirty beam
size.  A robustness of zero gave the best compromise between noise and
beam size.  The resulting RMS value was $2.7 \times 10^{-4}\;$Jy/Beam
and we cleaned our maps to a depth of 1-$\sigma$.  Images of HI
intensity, velocity field, and velocity dispersion were obtained with
XMOM in AIPS after blanking unnecessary channels with the task BLANK.

The velocity-integrated HI distribution is presented in
Fig.~\ref{HIint}.  The eastern spiral arm has a clear double ridge in
HI where it joins to the bar, while the inner arm fades farther out.
The western arm is also double, possibly even triple, but in this case
the innermost arm is that most easily traced to the outer disk.  As
shown in Fig.~7 of JvM95, the HI distribution extends only slightly
farther out than the light; there is very little neutral gas beyond a
deprojected radius of 400\arcsec, while we already reported above that
$R_{23.5}=348\arcsec$ in the I-band.  Furthermore, the outer
distribution is asymmetric, with a single spiral arm extending to the
NW.

\begin{figure}[t]
\centering
\includegraphics[angle=0,scale=0.6]{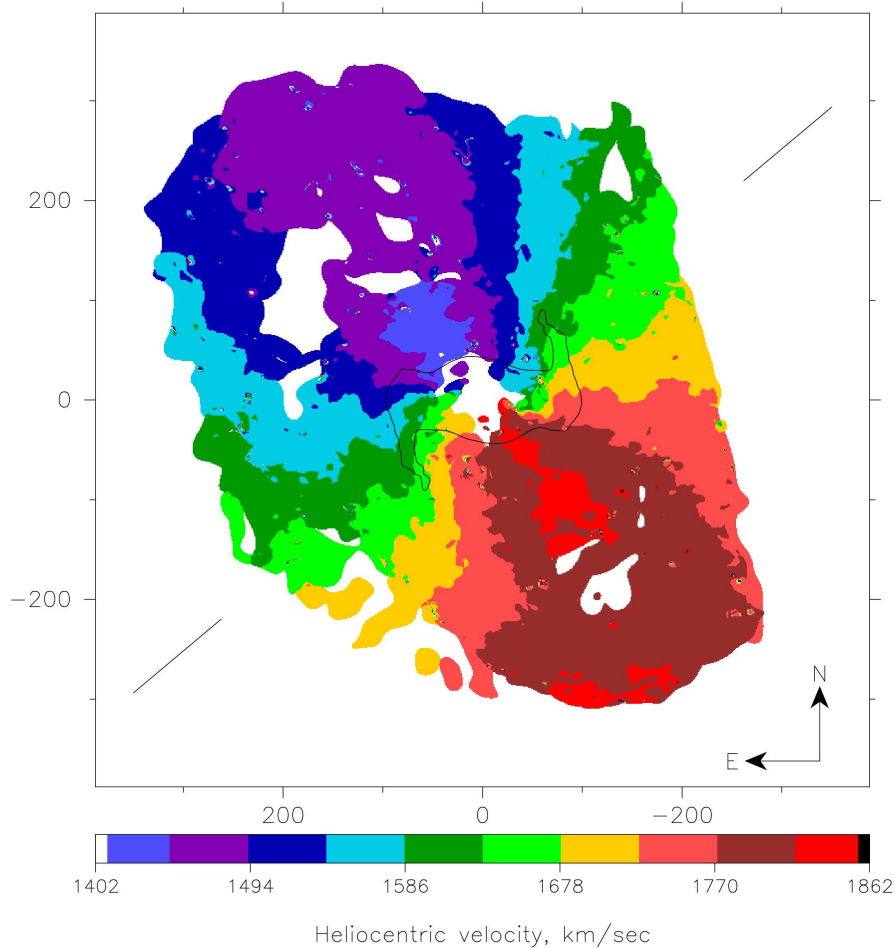}
\caption{\footnotesize The velocity map of NGC 1365 from observations
of the 21-cm emission line.  The color map has been binned in the same
10 colors used in Fig.~\ref{velfig} to outline velocity contours.}
\label{velHI}
\end{figure}

The velocity field mapped in the 21cm line, presented in
Fig.~\ref{velHI}, is nowhere characterized by a simple planar,
near-circular flow pattern.  The strong bar and spiral arms produce
non-axisymmetric distortions to the flow pattern that are revealed by
kinks in the isovelocity contours.  Also a mild warp is suggested by
the bending of the isovelocity contours towards the north in the
north-east side and towards the south in the south-west side of the
galaxy.  All of these factors conspire to make the rotation curve of
NGC 1365 very difficult to determine.  JvM95 concluded that the outer
disk was strongly warped, leading them to report a steeply declining
rotation curve.

\section{Asymmetries}
\label{asymmsec}
While asymmetries of all kinds are evident at larger radii, the inner
parts of NGC~1365 are at once remarkably symmetric in the I-band light
(Fig.~\ref{ibandfig}), and strikingly asymmetric in the position of
the dust lane and gas kinematics.  The \ha\ velocities in the bar
region also reveal a number of patches of emission with strongly
anomalous velocities when compared with the mean flow.  These facts
point to some kind of on-going disturbance to the gas flow in the
inner galaxy that has little effect on the light.

The galaxy is a member of the Fornax cluster; its projected position
is near the cluster center, while its systemic velocity differs from
the cluster mean by about 200 km/s \citep{Madore}.  Thus the outer
parts of NGC~1365 could be affected by the tidal field of the cluster
and/or ram-pressure of the intra-cluster gas.  However, tidal forces
are unlikely to be strong enough to affect the inner parts of this
massive galaxy, where asymmetries must have a different origin.

One possibility that NGC~1365 is in the advanced stages of a minor
merger.  However, the absence of noticeable disturbance to the stellar
bar requires that the infalling dwarf had a low enough density to have
been tidally disrupted before reaching the bar region.  Another
possible explanation might be that a stream of gas has fallen in,
perhaps from a tidally disrupted cloud or gas-rich dwarf galaxy.  Such
a stream, which would have to be more substantial than that detected
in NGC~6946 \citep{KS93,BHOFS} or the high-velocity clouds of the
Milky Way \citep{wvw97}, may have passed to the west of the center and
ahead of the leading side of the bar.  The ram pressure of this gas
acting on the gas in the galaxy mid-plane might have removed a good
fraction of the upstream gas ahead of the original position of the
shock, and allowing the shock to advance to its observed position.
This scenario is not without its problems, however; most notably, the
HI distribution is not noticeably depleted and the velocity pattern is
not strongly disturbed where the gas stream is supposed to have
punched through the mid-plane.

Whatever their origin, the peculiarities of the velocity field have
proved a major obstacle to the objectives of this study.

\section{Rotation Curve}
\label{rotcursec}
We first determine the center, systemic velocity, position angle and
inclination from the \ha\ velocity map using the $\chi^{2}$
minimization technique proposed by \citet{BS03}.  Their method uses
the entire kinematic map (excluding the sparse gas in bar region), in
order to estimate these projection parameters.  We find that the
center lies only a few arcsec from the reported position of the
nucleus \citep{Lindslit96}, and adopt their quoted position of the
nucleus as the kinematic center for the rest of our analysis.  The
fitted inclination of $41^\circ$ and PA of $220^\circ$ agree well with
the values obtained from the AIPS routine GAL, while the fitted
heliocentric velocity of 1631.5 for the \ha\ map and 1632 km/s for the
HI map are in excellent agreement with the value of 1632 km/s reported
by JvM95.

Applying a similar \chisq\ minimization to the I-band image yielded
the same PA but a higher inclination: $i=52^\circ$, which is in
tolerable agreement with the value $i=55^\circ$ estimated by
\citet{Lindblad78} also from photometric isophotes (see JvM95 and
references therein).  The large discrepancy between the kinematic and
photometric inclinations is probably due to the strong spiral features
that happen to lie near the projected major axis
\citep[see][]{palunas00, BS03}.  Since the spiral arms appear to bias
the photometric inclination more, we adopt the kinematic inclination
of $i=41^\circ$, consistent with most other work.  For completeness,
we have also followed our analysis through using the photometric
inclination, but find (\S~\ref{comparsect}) the kinematic inclination
leads to superior fits.

\begin{figure}[t]
\centering
\includegraphics[angle=270,scale=0.5]{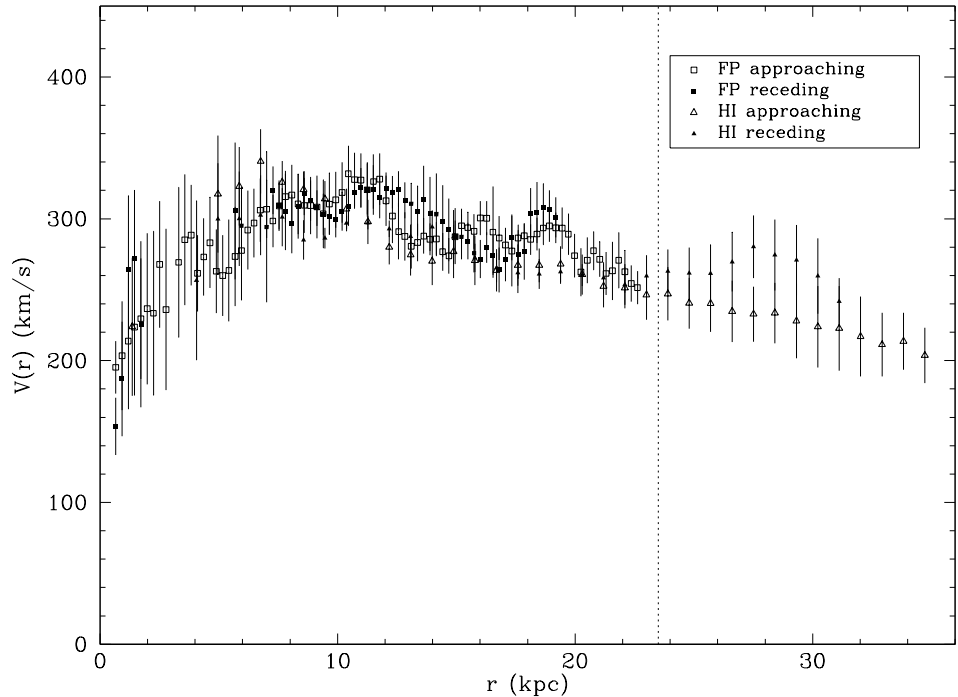}
\caption{\footnotesize Rotation curve from 21-cm and \ha\
observations. V(r) was fitted for receding and approaching sides
independently.  Annuli for Fabry-Perot (FP) are 3 arcsec wide and HI
map is 10 arcsec.  Data outside r=23.5 kpc, marked by the vertical
dotted line, has been ignored from our dark matter halo modeling.}
\label{RC_4_obs}
\end{figure}

In order to re-estimate the rotation curve from the HI velocity map,
we excluded gas beyond a deprojected radius of 255\arcsec\ (23 \kpc)
from the center.  Although the data extend 133\arcsec\ (12 \kpc)
farther, pronounced asymmetries in these outer parts suggest the gas
layer is disturbed and may not be in simple rotational balance.  While
we cannot exclude the possibility that the gas layer in our selected
inner region is warped, we here adopt the simpler assumption that the
gas in the inner 255\arcsec\ is everywhere flowing in a single plane.

With the PA, inclination and systemic velocity fixed, we fitted for
the circular velocity in annuli of 3 arcsec and 10 arcsec wide for the
\ha\ and HI data respectively.  We fitted the approaching and receding
sides separately, producing four separate estimates of the rotation
curve, which we present in Fig.~\ref{RC_4_obs}. Squares (triangles)
represent fits to the Fabry-Perot (HI) observations, and we use open
symbols for the approaching side of the galaxy and filled ones for the
receding side.  The uncertainties in the bar region are large because
we here fit simple rotational motion to a flow pattern that is
manifestly non-circular; we improve on this below.  Fits with standard
deviations greater than 60 km/s were discarded.  In the inner disk,
where we expect the galaxy to be flat, we find the four fitted
velocities generally agree within the estimated uncertainties.
However, there are substantial differences between the receding and
approaching sides in the HI data beyond the vertical dotted line drawn
at $R=23\;$kpc.  Accordingly, we exclude data beyond 23~kpc (the
region excluded for finding PA, inclination), for fitting our dark
matter halo (\S\ref{halomodelsec}).

The rotation speed declines significantly outside 10~kpc, although not
as steeply as that derived by JvM95.  A decline within the visible
disk is not uncommon for galaxies of this luminosity
\citep{CvG91,noordermeer07}.

\section{Mass models}
The first stage of our modeling procedure is to build a family of
axisymmetric mass models with differing mass disks that, when combined
with a halo, yield our adopted axisymmetric rotation curve, at least
in the outer parts.

\subsection{Disk}
We compute the disk contribution to the central attraction by first
assuming a fixed M/L ratio ($\Upsilon_I$) for all the luminous matter,
after rectifying the I-band image to face-on and applying a global
correction for extinction.  We used the inclination and PA obtained
from the kinematic maps to perform the deprojection.

A constant $\Upsilon_I$ is a good assumption provided that the stellar
population is homogeneous and dust obscuration is negligible.  Most
variations in the V-I color visible in Fig.~\ref{dustfig} appear to be
due to dust, which affects the V-band surface brightness much more
than the I-band.  While Fig.~\ref{ibandfig} shows that dust still
diminishes the I-band surface brightness, especially in the bar
region, we have not attempted to correct for the patchiness of the
dust.

\begin{figure}[t]
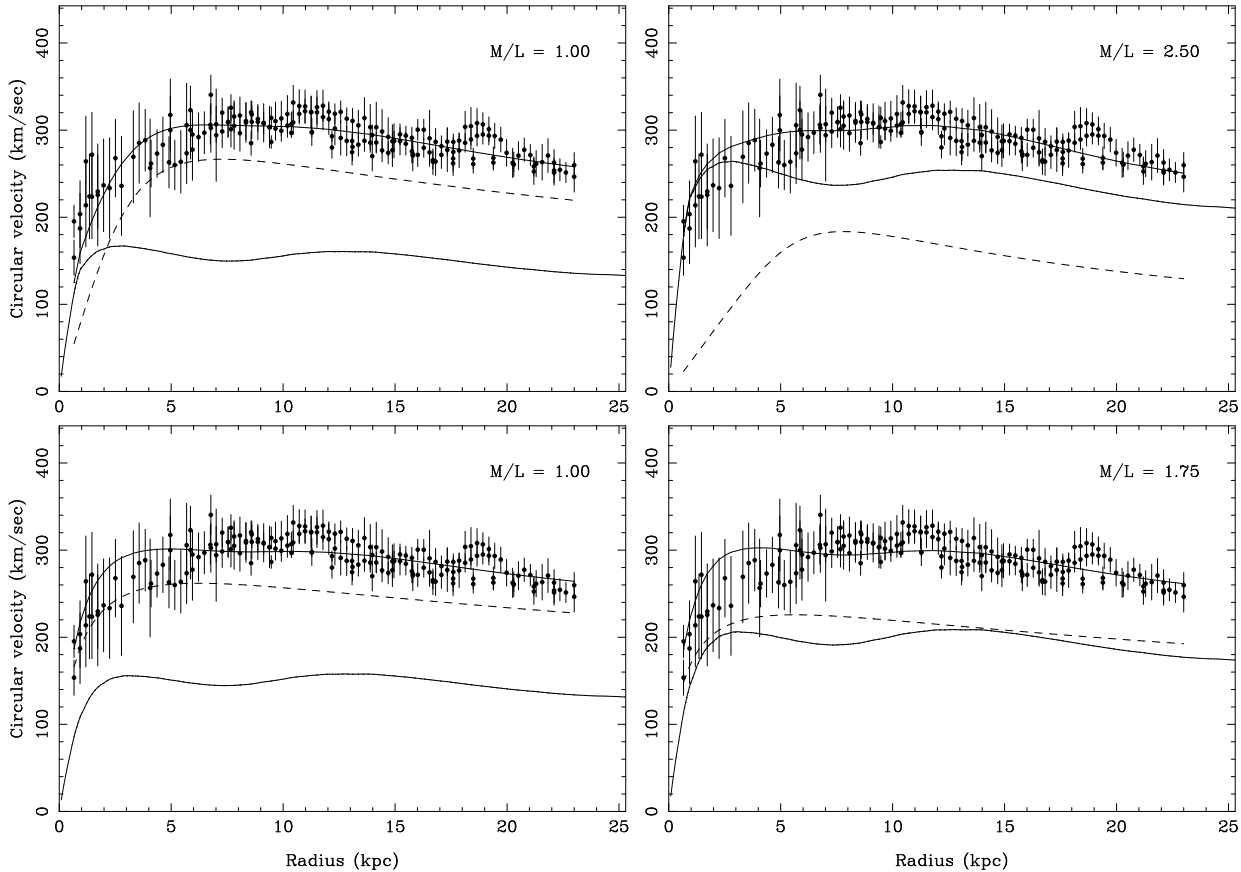

\centering

\includegraphics[angle=270,scale=0.35,viewport=0 0 450 670,clip]{ml100b1075.ps}
\includegraphics[angle=270,scale=0.35,viewport=0 25 450 670,clip]{ml250b1075.ps}\\

\includegraphics[angle=270,scale=0.35,viewport=0 0 500 670,clip]{ml100NFW.ps}
\includegraphics[angle=270,scale=0.35,viewport=0 25 500 670,clip]{ml175NFW.ps}\\

\caption{\footnotesize Mass models for the rotation curve.
\textit{Points and error bars:} Observed rotation curve in both HI and
\ha.  \textit{Uppermost solid line:} Best-fit total rotation curve.
\textit{Dotted line:} Contribution of the best-fit halo.
\textit{Lower solid line:} Contribution of the stellar disk to the
total rotation curve. The left panels have disk $\Upsilon_I = 1.0$ and
the right panels have disk $\Upsilon_I = 2.50$ and $\Upsilon_I =
1.75$. The top panel use pseudo-isothermal halos with bulge
b1075. Bottom panels use NFW halos.}
\label{rotcurvefig}
\end{figure}

As we need the in-plane forces for both the strongly non-axisymmetric
light distribution, and for an azimuthally averaged light profile, we
adopt the same numerical procedure for both.  We assume a finite
thickness for the disk with the usual mass profile normal to the plane
of the galaxy: $\rho(z) \varpropto {\rm sech}^2(z/2h_z)$ with
$h_z=0.5\;$\kpc\ consistent with observations of edge-on galaxies for
our measured disk scale length $r_d = 6.5\;$\kpc\ \citep{KvdKdG}.

Forces from the visible matter scale linearly with the adopted M/L
ratio; we explore a discrete set of values for $\Upsilon_I$ ranging
from 0.50 to 3.75 in steps of 0.25.  Axisymmetric rotation curves from
the visible matter for three different $\Upsilon_I$ values are shown
by the solid lines in Fig.~\ref{rotcurvefig}.

We have not explicitly determined the contributions from the mass of
the neutral and molecular gas to the rotation curve.  Most of the
neutral gas, $\sim 1.7\times10^{10} M_{\sun}$ after helium correction,
is outside the bar whereas all the molecular gas is inside the bar and
nucleus, $\sim 1.7\times10^{10} M_{\sun}$ of which $5.4\times10^{9}
M_{\sun}$ lies in the central 2.0 kpc \citep[see][]{sandqvist}.
Altogether gas accounts for about 10\% of the total mass, as measured
from the HI rotation curve (JvM95).  The combined distribution of the
two phases has a density profile that resembles that of the light and
therefore can be taken into account by simply associating the stellar
mass-to-light ratio to be roughly 90\% of the fitted value.  While
this approximation neglects the different thicknesses of the star and
gas layers, the total rotation curve will be very little affected by
taking into account the marginally stronger central attraction from
the small fraction of mass in the gas.  \Ignore{A more detailed
treatment would not be meaningful because of the low-resolution of the
HI data and the lack of an extensive map of the molecular gas
distribution.}

\subsection{Bulge}
\label{bulge}
The assumptions of constant $\Upsilon_I$ and disk thickness at all
radii are clearly the simplest we could adopt, but may not be correct,
especially in the center.  While the light distribution in the inner
bar region is somewhat rounder than that in the outer bar, it does not
really suggest a spheroidal bulge because deprojection, which stretches
the image along the minor axis, still leaves this component roughly
aligned with the bar.  However, it is possible that the inner
bulge-like feature is thicker than the disk and bar and it may also
have a higher $\Upsilon_I$ than the rest of the disk.

To allow for a possible enhanced $\Upsilon$ in the inner disk, we
consider some simple variants of our constant $\Upsilon_I$ models.
For $r \leq r_b$, we scale the $\Upsilon_I$ by the function
\begin{equation}
F(r)=1 + f\times\left( 1-\frac{r}{r_b}\right) \left(1+\frac{r}{r_b}\right)^2,
\end{equation}
where $r_b$ is the radius of the bulge and $f$ is a constant.  We have
adopted three separate values of $r_b$ and of $f$, and label the five
bulge models we study here as b1025, b1075, b2050, b2075, b3075.  The
first numerical digit denotes $r_b$ in kpc and the last 3 digits are
100 times the value of $f$; \ie\ b2050 denotes a model bulge with
$r_b=2\;$kpc and $f=0.50$.

This approach assumes that the possible bulge is as thin as the outer
disk, which may not be true.  A thicker component would exert slightly
weaker central forces, which will result in our fits preferring a
slightly lower value of $f$ than the population has in reality.
However, we do not need to know the relative flattening of this mass
component; our procedure simply allows for the possibility that it may
give rise to some extra central attraction.

As found by \citet{JCA97} for their H-band image, the inner light
profile of our I-band image is extended at different position angle
from that of the main bar.  Jungwiert \etal\ and \citet{LSKP} suggest
this is evidence of decoupled nuclear bar although an HST NICMOS
image resolves this feature into a nuclear spiral
\citep{Emsellem01,Erwin04}.  Whatever the correct interpretation, this
distortion in the inner light distribution is inside the region we
exclude from our fits to the observed velocities, and does not affect
our conclusions.

\subsection{Halo}
\label{halomodelsec}
The total central attraction should account for the rotational
balance, and generally requires a dark halo. We have adopted 2
different dark halo forms: a generalized pseudo-isothermal halo and
the NFW density profile \citep{NFWpaper}.

In order to fit the declining rotation curve, we
adopt a slightly generalized pseudo-isothermal dark matter halo
density profile of the form proposed by \citet{albada3198}:
\begin{equation}
\rho(r)=\rho_o\frac{1}{1+(r/r_c)^k}.
\end{equation}
The exponent $k$ is treated as an independent parameter that allows
the halo circular speed to decline if $k>2$.  Since large
values of $k$ lead to a halo rotation curve with a sharp, narrow
peak, we apply the additional constraint $k\leq5$.  

NFW \citep{NFWpaper} dark halo profiles have the form
\begin{equation}
\rho(r)=\frac{\rho_sr_s^3}{r(r+r_s)^2}.
\end{equation}

For both density profiles we fit the total rotation curve by
minimizing $\chi^2$ to find the best fit halo to be combined with each
adopted disk $\Upsilon_I$ and bulge model.  The results are given in
Table \ref{halotab} and Table \ref{halotabNFW}, with the values of $c$
and $V_{200}$ being computed for a Hubble constant $H_0=
70\;$\kms~Mpc$^{-1}$.  No bulge modification was applied to the NFW
halo since extra bulge mass combined with the cuspy halo would have
complicated the fits to the inner rotation curve.
Figure~\ref{rotcurvefig} shows possible fits to the data with
different $\Upsilon_I$ exemplifying the disk-halo degeneracy for both
dark matter halo profiles adopted.  The pseudo-isothermal profile
plotted uses a bulge model b1075.

Both our adopted halo functions assume spherical symmetry, whereas we
need assume only that the halo is axisymmetric in the disk plane.  The
flow pattern in our simulations depends on $\Upsilon_I$ and the
corresponding halo attraction; the possible flattening of the halo
affects only the interpretation of the halo attraction in terms of a
mass profile.  As a flattened halo gives rise to stronger central
attraction in the mid-plane for the same interior mass \citep[\eg\
Fig.~2-12][]{BT87}, the simplest assumption of spherical symmetry
allows a more massive halo than would be required if the halo were
significantly oblate.

\section{Gas Dynamics simulations}
We here give a brief summary of our procedure; see \citet{wein2} for a
more detailed description.  As usual, we neglect self-gravity of the
gas, which would add considerably to the computational complexity of
our many simulations.  This simplifying assumption is justified
because the gas density in the bar, where we are most interested in
comparing with the observed velocities, is a small fraction of the
mass in stars.

We compute the gas flow using the flux splitting code written and
tested by \citet{codepaper} and kindly provided by Athanassoula.
\citet{P07} has shown that results from the 2-D code we use and the
3-D SPH code used in her work agree very well for the same mass model,
which is reassuring.

Since the I-band light has excellent 2-fold rotation symmetry
(Fig.~\ref{ibandfig}), we force bi-symmetry in the simulations and
employ an active grid covering only one half of the galaxy.  In order
to avoid initial transients resulting from non-equilibrium starting
conditions, we begin our simulations in an axisymmetric potential, and
gradually change to the full non-axisymmetric barred potential over
the first 2.5 Gyr of evolution.  Careful inspection of the simulations
revealed that the gas does not settle to a quasi-steady flow until
several bar rotations (3 Gyr or from 6 to 15 rotations for
$\Omega_p=12$ and 32 respectively) are completed.  At this time, we
take a snapshot of the simulation and compare it to the data.  We
stress that the final state of the gas doesn't depend on the growing
time, but the bar growth rate does affect how quickly the gas flow
settles down.

We experimented with different sound speeds and grid resolutions,
which generally yielded reproducible results except in the central few
resolution elements.  We found that flows at the highest resolution
(grid cell size 128~pc) did not ever settle to quasi-steady patterns
for the most massive disks, but did do so for a cell size of 300~pc;
this change in behavior is caused by the stabilizing influence of
greater numerical viscosity on the coarser grid.  We therefore work
with the results from simulations with a grid cell size of 300~pc for
all disk masses.

Modeling the ISM with a simple fluid code is clearly an approximation.
An isothermal equation of state with sound speeds in the range 6 to
$10\;$\kms\ is reasonably appropriate for the warm neutral and warm
ionized components, which have temperatures of $\sim 10^4\;$K, and we
attempt to fit our models to the observed kinematics of the ionized
component.  The neutral HI component of the Milky Way \citep{GKT} and
other galaxies \citep[\eg][]{Kamphuis93} is also observed to have
intrinsic line widths in this range.  Furthermore, dense clouds are
observed also to maintain a velocity dispersion of $\sim 8\;$\kms\
\citep{SB89}.  \citet{EG97} showed that the gas flow pattern in their
simulations changed materially when they increased the sound speed to
25~\kms; however, such a large sound speed is not representative of
any dynamically significant component of the ISM.  We therefore
confine our tests of the gas parameters to realistic ranges, and find
the significant properties of the flow are almost independent of the
exact values adopted.

The non-circular flow pattern produced by a bar depends both on the
mass of the bar and its pattern speed or rotation rate \citep{rob79}.
For the photometric disk model described above, the bar mass scales
directly with the adopted $\Upsilon_I$, while the total central
attraction includes the contribution from the adopted halo, which is
assumed to be axisymmetric.  For each adopted $\Upsilon_I$ ratio, we
compute the gas flow in the disk plane using 2-D gas dynamics
simulations, for a number of different bar pattern speeds, $\Omega_p$.
We project the resulting flow patterns as the galaxy is observed, and
compare with Fabry-Perot velocity maps using a \chisq\ analysis to
find the best match to the observations, in order to estimate the disk
$\Upsilon_I$ and $\Omega_p$.

We explored pattern speeds in the range $12 \leq \Omega_p \leq
32\;$\kms~kpc$^{-1}$, corresponding to corotation radii in the range
$2.0r_{B}\ga r_L \ga 0.9r_{B}$.  In all we ran 154 simulations to
cover the grid in $\Omega_p$ and $\Upsilon_I$ for each of the
pseudo-isothermal and NFW halo models.

\subsection{Comparison between models and data}
\label{comparsect}
Before attempting any comparison between simulations and data, it is
necessary to smooth the simulations to the atmospheric seeing during
the observations and also by the largest kernel used to produce the
velocity map (see \S\ref{fabrysec}), which requires a smoothing
kernel of FWHM~$=4\arcsec$.  \citet{Kamphuis93} reports that the
velocity dispersion of HI gas rises inwardly for a number of galaxies
to values that exceed 10~\kms\ in the bright inner parts of massive
galaxies.  Since the velocity in each pixel may reflect that of an
individual HII region, we add 12~\kms\ in quadrature to the formal
error estimates of the \ha\ velocities to allow for a possible
peculiar velocity relative to the mean flow.

Since our simulations assume a gas with a finite sound speed, the full
non-linear velocity distribution at every point takes into account our
adopted velocity spread.  No further correction, \eg\ for asymmetric
drift, is therefore needed.

We compare the projected snapshot from the simulation using the
standard goodness of fit estimator
\begin{equation}
\chisq = {1 \over N} \sum_{i = 1}^N z_i^2,
\label{chisqeq}
\end{equation}
where the summation is over all $N$ pixels in the region selected for
comparison, and $z_i$ is the usual difference between the observed
velocity $V_{i,{\rm obs}}$ and that predicted from the simulation
$V_{i,{\rm mod}}$, weighted by the uncertainty $\sigma_i$; viz.
\begin{equation}
z_i = {V_{i,{\rm obs}} - V_{i,{\rm mod}} \over \sigma_i}.
\end{equation}
We include 2280 pixels in the fit, but they are not all independent
because of seeing and smoothing.  Furthermore, the quantity \chisq\ is
regularized somewhat by adding 12~\kms\ in quadrature to the errors.
Thus it is not a formal estimator of confidence intervals, but can be
used to compare relative goodness of fit.

Our mass model is bi-symmetric, in line with the light in this galaxy,
but the velocity data are not.  Our simulations are therefore unable
to fit both sides of the bar simultaneously.\footnote{Since the I-band
light distribution is almost perfectly bisymmetric (Fig.~1), relaxing
the assumption of bi-symmetry had a negligible effect on the model
predictions.}  Since the dust lane and steep velocity gradients are
quite uncharacteristically almost outside the bar on the west side, we
confine our fits to the east side, where we suspect that the flow
pattern is less disturbed.

We further restrict the region from which we evaluate \chisq\ to only
those pixels in the area between the inner ellipse and the outer
rounded rectangle shown in Fig.~\ref{velfig}, because our models are
most sensitive to the parameters we have varied in this region.  The
inner ellipse has a semimajor axis of 35\arcsec and ellipticity $0.5$,
while the semi-axes of the outer curve are 80\arcsec \& 42\arcsec.  We
exclude the center for several reasons: the observed velocities are
degraded somewhat by beam smearing, our assumptions of a thin disk are
most likely violated in the center (see section \ref{bulge}), and the
predicted gas velocities vary with grid resolution in this region.  We
exclude data in the spiral arms because the gas flow outside the bar
never settles to a steady pattern in our simulations, and some studies
\citep[\eg][]{sellwood88} suggest that these features evolve over
time.

The value of \chisq\ varies substantially as the two parameters are
varied, but the minimum remains large, even when we restrict the
region of comparison with the data to the eastern half of the bar.
Maps of the residuals, shown below, reveal a number of isolated
regions with substantially discrepant velocities that coincide with
the regions labeled in Fig.~\ref{velfig}.  In order to check whether
the position of the minimum was being displaced by the pixels with
large anomalous velocities, we employed Tukey's biweight estimator
\citep{recipes}
\begin{equation}
\chi_{\rm bw}^2 = {1 \over N} \sum_i \cases{
z_i^2 - z_i^4/c^2 + z_i^6/(3c^4) & $|z_i|<c$, \cr
c^2/3 & otherwise, \cr}
\label{biwteqn}
\end{equation}
with the recommended value for the constant $c=6$.  In fact, the
position remained unchanged for both the isothermal and NFW halos, and
we revert for the remainder of the paper to the usual \chisq\
statistic defined by eq.~(\ref{chisqeq}).

\begin{figure}[t]
\centering
\includegraphics[angle=270,scale=0.63]{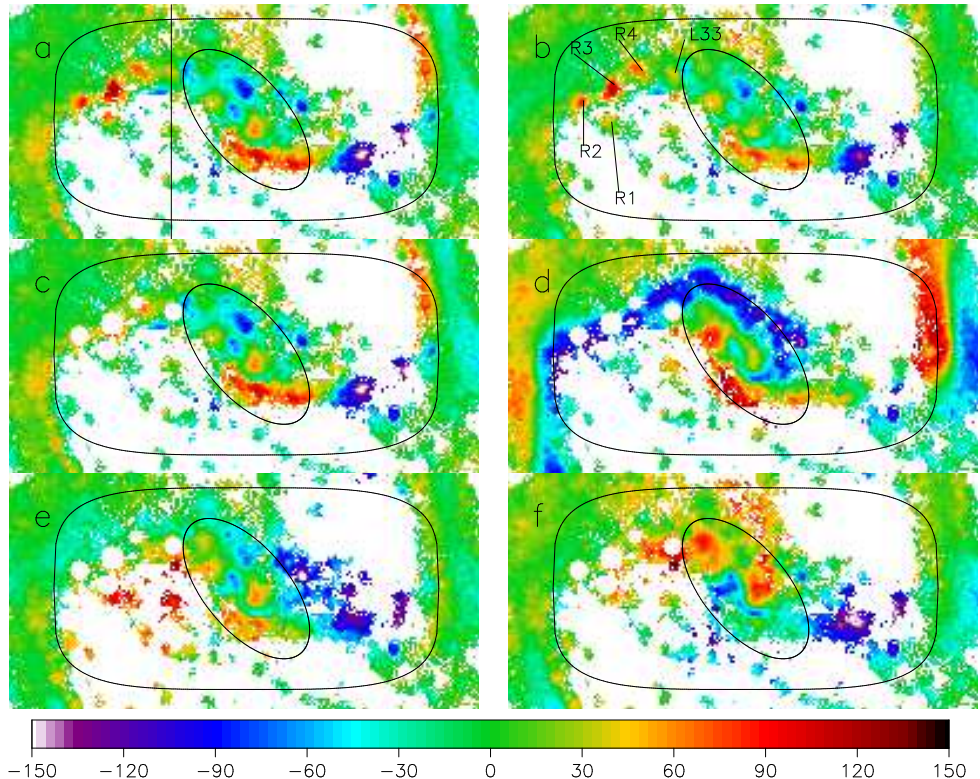}
\caption{\footnotesize (a) \& (b) Residual maps for respectively the
best-fit pseudo-isothermal and NFW halo models.  The velocities of the
anomalous HII regions, identified in (b), are shown in these two
panels only.  The N-S line in (a) is described in \S\ref{slitsec}.
(c)-(f) 4 characteristic b1075 models, with the anomalous HII regions
masked out: \textit{c.)}\ $\Upsilon_I=2.50$, $\Omega_p=$24;
\textit{d.)}\ 1.0,24; \textit{e.)}\ 2.50,16; \textit{f.)}\ 1.0,16.}
\label{chismapfig}
\end{figure}

The minimum of the reduced \chisq\ is unreasonably high: $\chisq =
2.7$ for the best isothermal halo and $\chisq = 2.4$ for the NFW halo.
It is evident from the residual maps (Figure~\ref{chismapfig}a \& b)
that a large part of the mismatch from the model comes from the five
anomalous regions identified already in Fig.~\ref{velfig}.  We
therefore masked each of these regions with circular mask of radius 4
or 5 pixels, which decreased the minimum of \chisq\ to 1.6 for the
isothermal models and 1.5 for then NFW halo models.  While these
values are still on the high side, we should not expect $\chi^2=1$
since the fluid model is idealized.  We therefore consider we have
satisfactory models for most of the flow pattern in the eastern half
of the bar.

Since we approximate the ISM as an isothermal gas, the shock in the
simulations must be as close as the code can resolve to a
discontinuity, whereas it is possible that gas velocities in NGC~1365
vary more gradually.  We therefore experimented with additional
smoothing of the simulations above the FWHM~$=4\arcsec$ that matches
the resolution of our data.  The value of \chisq\ decreases
considerably to $\chisq=1.3$ when we smooth the simulations by
10\arcsec, although the position of the minimum did not change.  At
least some of this improvement appears to result from a better match
to the velocity gradient across the shock.  We have not included this
extra smoothing in the fits below, however, because it degrades our
resolution everywhere and reduces the curvature of the \chisq\
surface.

\begin{figure}[t]
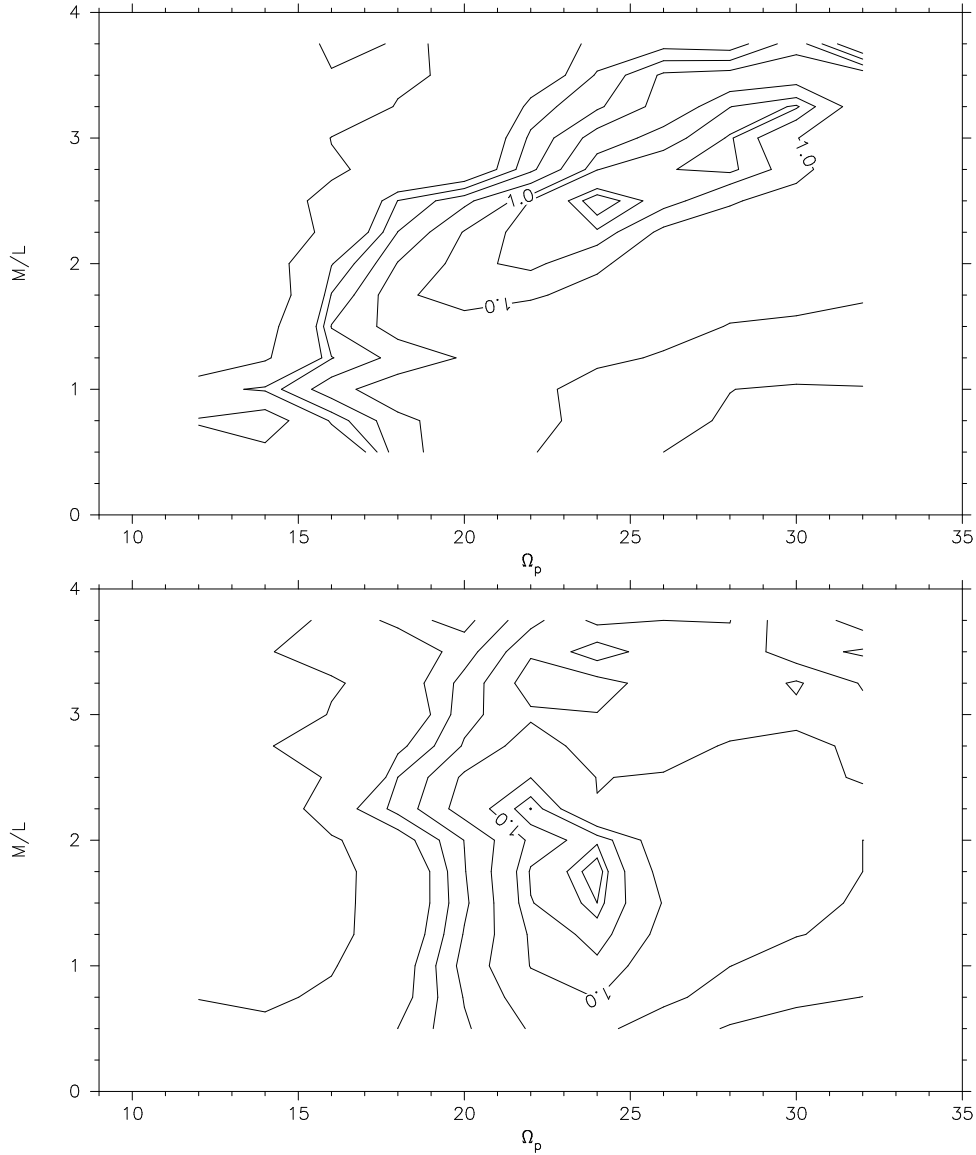

\centering
\includegraphics[angle=270,scale=0.5]{chisqconb1075.ps} \\
\includegraphics[angle=270,scale=0.5]{chisqconNFW.ps}
\caption{\footnotesize \textit{Upper:} The \chisq\ surface for the b1075
pseudo-isothermal model.  The minimum value (1.6) lies at
$\Upsilon_I=2.50$ and $\Omega_p=24$.  Contours are drawn at
$\Delta\chisq/N 0.1$, 0.2, 0.5, 1, 2, 3, 4, 5 \& 10 above the minimum.
\textit{Lower:} The same for the NFW halo model.  The minimum is 1.5
for model with $\Upsilon_I=1.75$ and $\Omega_p$=24.}
\label{chiscontfigL}
\end{figure}

\subsection{Best fit models}
Figure~\ref{chiscontfigL} shows contours of \chisq\ in the space of
the two principal parameters: $\Upsilon_I$ \& $\Omega_p$.  The
quantity contoured is the value from eq.~(\ref{chisqeq}) using pixels
from the east side of the bar only with the 5 anomalous regions masked
out.

\begin{figure}[t]
\centering
\includegraphics[angle=270,scale=0.63]{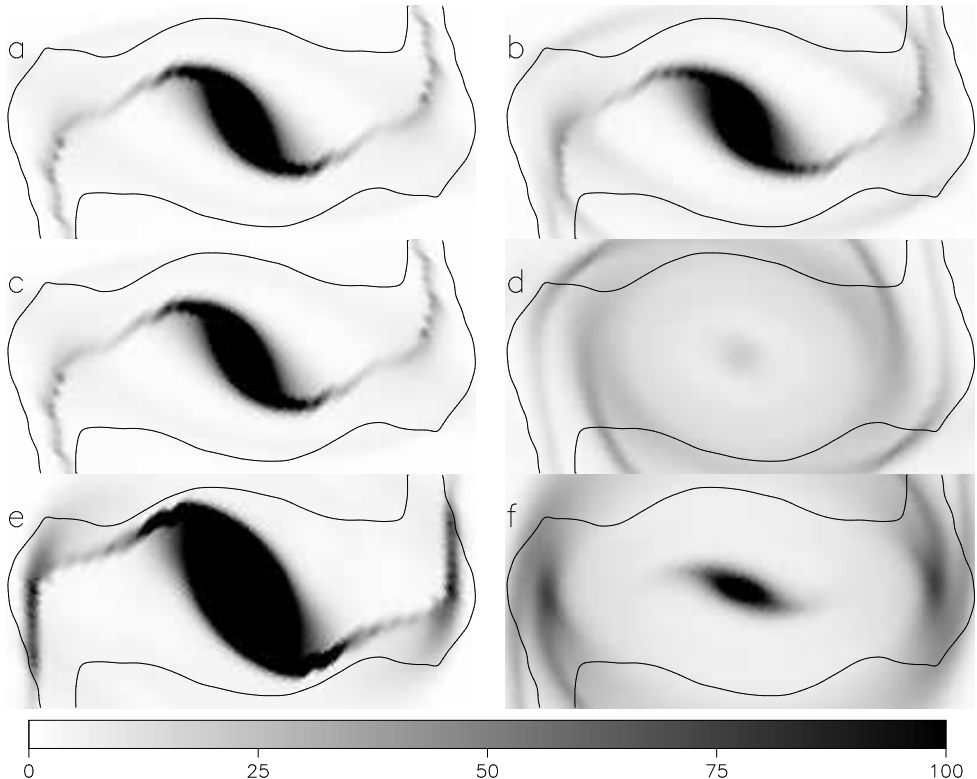}
\caption{\footnotesize Gas surface density in the same models shown in
Fig.~\ref{chismapfig}, panels (a) \& (c) are identical, therefore.
Panels (c) thru (f) show that the density response to bar forcing
changes dramatically as the disk mass and pattern speed are varied.}
\label{gasdensfig}
\end{figure}

For the pseudo-isothermal halos, models with low pattern speed and
high $\Upsilon_I$ are very strongly disfavored and the \chisq\
function also rises strongly, but somewhat less steeply towards higher
pattern speeds and low $\Upsilon_I$.  We show the residual map for the
best fit parameters and, for comparison, three other cases for two
different values of $\Omega_p$ and $\Upsilon_I$ in
Figure~\ref{chismapfig}(c)-(f).

Figure~\ref{gasdensfig} shows the gas density response in the inner
regions of the same set of models.  Shocks are regions of locally high
gas density along the bar that coincide with steep velocity gradients
in the flow.  The shocks in the best fit models (top row) are
displaced towards the leading edge of the bar, as usual, and all-but
disappear in the low-mass disks, panels (d) \& (f).  The shocks in
panel (e) are too far towards the bar's leading edge, and the inner
oval is too large, because the $x_2$ family that is responsible for
this behavior \citep{SW93} grows in spatial extent as the pattern
speed is reduced.  Although the position of the dust lane in the west
of the bar is indeed shifted towards the leading edge, it does not
have the shape predicted by low pattern speed models.

It is difficult make a more quantitative comparison between the gas
density in the models and that observed.  We do not have sufficient
information about the distribution of all the multiphase components of
the ISM; we do not have a complete CO map, and the spatial resolution
of the HI map from the VLA is too low to attempt a meaningful fit.

The best fit pattern speed for both the pseudo-isothermal and NFW
halos is for $\Omega_p = 24\;$\kms$\rm{kpc}^{-1}$ which places
corotation at $r_L \approx 1.23r_B$.  This value of $\Omega_p$ is in
good agreement with that obtained by \cite{Linlinatha} in their
bar-only model (after correcting their assumed distance to ours).

However, the minimum \chisq\ is at $\Upsilon_I=2.50$ for the
pseudo-isothermal halos, whereas the somewhat lower value $\Upsilon_I
= 1.75$ is preferred for NFW halo models.  The rotation curve
decompositions for these two models are show in the right hand panels
of Fig.~\ref{rotcurvefig}.  An estimate of the statistical uncertainty
in the parameters is indicated by the $\Delta\chisq = 1$ contour in
Fig.~\ref{chiscontfigL}.  Both results are in agreement with each
other within the uncertainties and favor fast bars and moderately
massive disks.

Finally, we reworked our entire analysis using the photometric
inclination of $i=52^\circ$ instead of the conventionally adopted
kinematic inclination of $i=41^\circ$ in the foregoing analysis.  This
required redetermining the gravitational potential of the differently
oriented and projected bar and disk, refitting for the halo parameters
for each adopted $\Upsilon_I$, and running a new grid of models over
the two free parameters.  The resulting fits were far worse, with the
minimum value of $\chisq \simeq 3.8$, even after masking out the HII
regions having anomalous velocities.  This test confirms that the
kinematically-determined inclination is the more appropriate.

\begin{figure}[t]
\centering
\includegraphics[angle=0,scale=0.5]{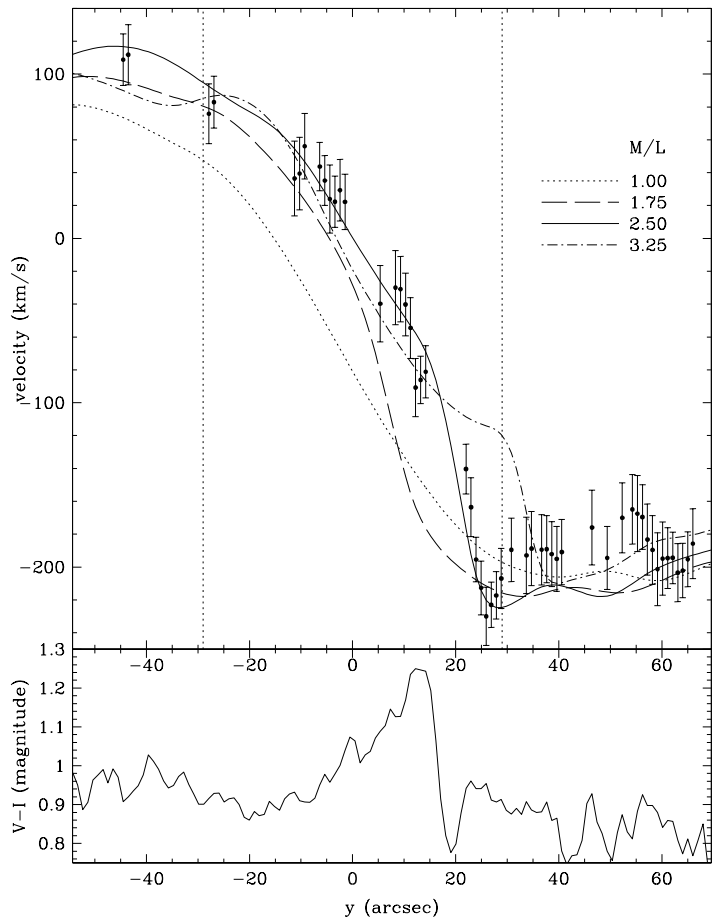}
\caption{\footnotesize Data points with error bars show the H$\alpha$
velocities along a pseudoslit passing perpendicular through the bar as
indicated in Fig.~\ref{chismapfig}(a). The two vertical dotted lines
mark the width of the bar. Different curves show the simulated
velocities in models with different $\Upsilon_I$ ratios for
$\Omega_p=24$ and the b1075 model.  Lower panel shows the color
profile along the same pseudoslit through our extinction map
(Fig.~\ref{dustfig}).}
\label{slitfigml}
\end{figure}

\begin{figure}[t]
\centering
\includegraphics[angle=0,scale=0.5]{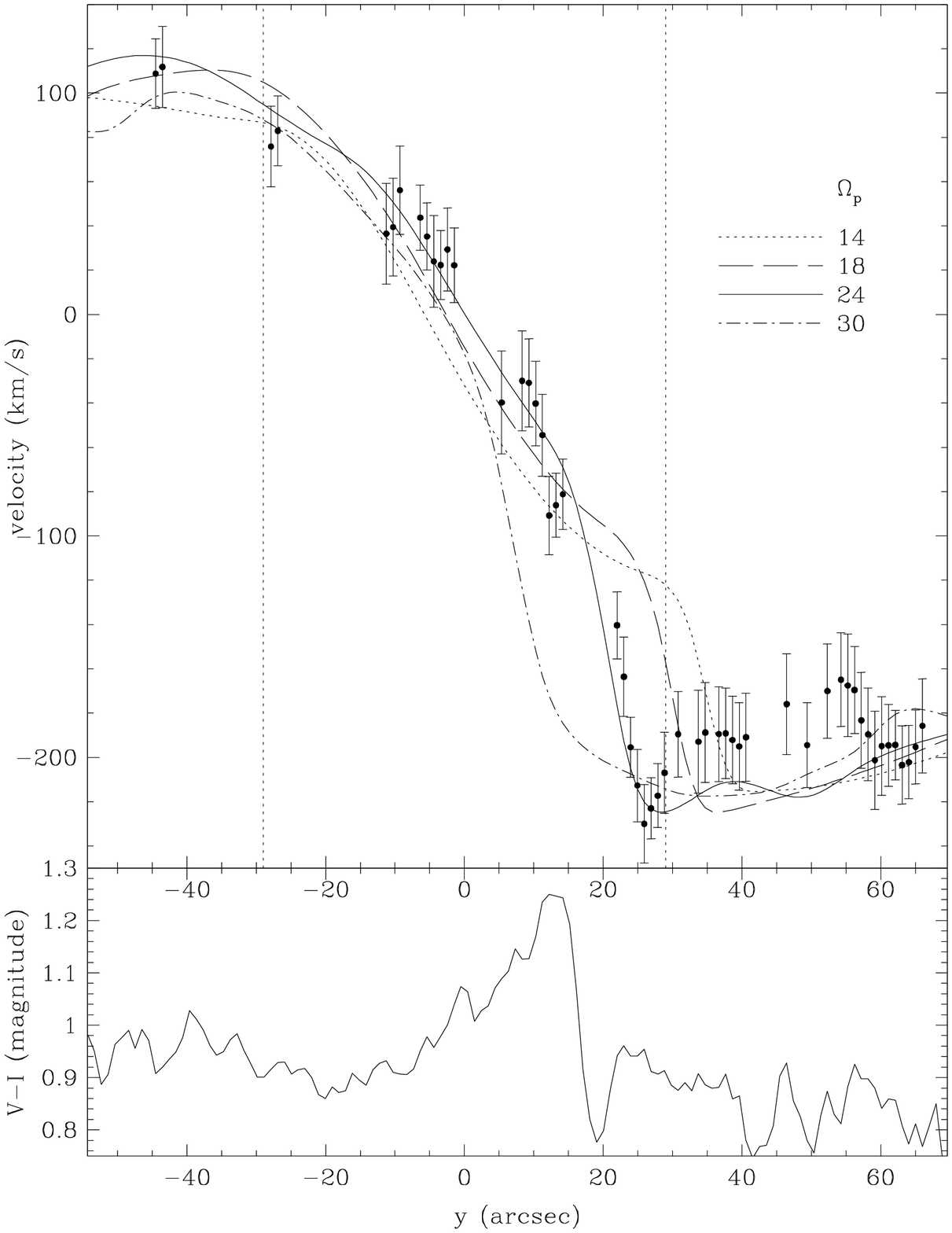}
\caption{\footnotesize As for Fig.~\ref{slitfigcr} but here the curves
show simulations having different $\Omega_p$ for fixed
$\Upsilon_I=2.50$.}
\label{slitfigcr}
\end{figure}

\subsection{Pseudo-slit cut through data}
\label{slitsec}
Shocks across dust lanes are readily identified as steep velocity
gradients. The data points in Figs.~\ref{slitfigml} and
\ref{slitfigcr} show the observed velocities along a pseudo-slit
placed perpendicular to the bar major axis as shown in
Fig.~\ref{chismapfig}(a); this cut across the bar has the most
extensive data.  The lower panel shows the V-I color along the same
line, which reveals that the projected velocity gradient of almost
$200\;$\kms\ is coincident with the prominent dust lane.  The vertical
dotted lines mark the bar width as presented in the bar isophote in
Figs.~\ref{ibandfig} and \ref{dustfig}.  The other lines in the upper
panel show velocities from our b1075 isothermal simulations for
different $\Upsilon_I$ with a fixed $\Omega_p=
24\;$\kms$\rm{kpc}^{-1}$ (Fig.~\ref{slitfigml}) and for different
$\Omega_p$ for a fixed $\Upsilon_I=2.50$ (Fig.~\ref{slitfigcr}).

These figures show that $\Upsilon_I = 1$ does not produce a strong
enough shock, and $\Omega_p \neq 24$ puts the shock/velocity gradient
in the wrong place.  In fact, the systematic dependences of the
velocity gradient on $\Upsilon$ and on $\Omega_p$ are quite similar to
those shown in the pseudo-slits for NGC 4123 \citep{wein2}, where the
effects of these physical parameters are explained.

\section{Discussion}
In the previous section, we argued that the underlying gas flow in the
eastern half of the bar can be modeled with a somewhat massive disk
embedded in one of two different halos.  Neither model makes a
compelling fit, however, and the properties of both best-fit halo
models are distinctly non-standard.  The outer profile of both our
fitted dark halo models already yields a declining circular speed by
the edge of the visible disk.  This finding may simply be a reflection
of a mis-estimated rotation curve, possibly due to a warp or tidal
disturbance.  However, in their attempt to fit a warp, JvM95 obtained
a more strongly declining rotation curve, which would require a still
less orthodox halo model.

\subsection{Disk mass}
Our principal finding is that the bar must be massive and rapidly
rotating in order for the underlying gas flow pattern to resemble that
observed.  We obtain improved fits with the isothermal model when we
increase the baryonic mass in the inner (bulge-like) region and also
when we adopt a halo function that has a high density near the center.
This fact suggests that the flow pattern requires a high density in
the center, but cannot determine whether the mass should be flattened
or spherical, baryonic or dark.

\subsection{Dynamical friction}
The disk contribution to the central attraction in our best fit model
with the NFW halo function, though large, is not fully maximal
(Fig.~\ref{rotcurvefig}), yet the bar is required to be fast.  This
model therefore presents the additional puzzle that dynamical friction
\citep{DS00} has not slowed the bar.  The uncertainty in $\Upsilon_I$
is great enough that more massive disks are acceptable, although such
models would further increase the uncomfortably high ratio of baryonic
to dark mass of the best fit model.  A lower value of $\Upsilon_I$ may
allow a more reasonable baryonic fraction, but exacerbates the puzzle
of the fast bar in a yet denser halo.

\subsection{Halo decompression}
The halo of our best-fit NFW model has $V_{200} = 111\;$\kms\ and a
concentration $c=61$.  These parameters, which are defined in
\citet{NFWpaper}, are not in line with the predictions of LCDM theory
\citep[\eg][]{BKSS01}; the concentration is very high and the
value of $V_{200}$ much lower than expected for a large galaxy.

\begin{figure}[t]
\centering
\includegraphics[angle=270,scale=0.5]{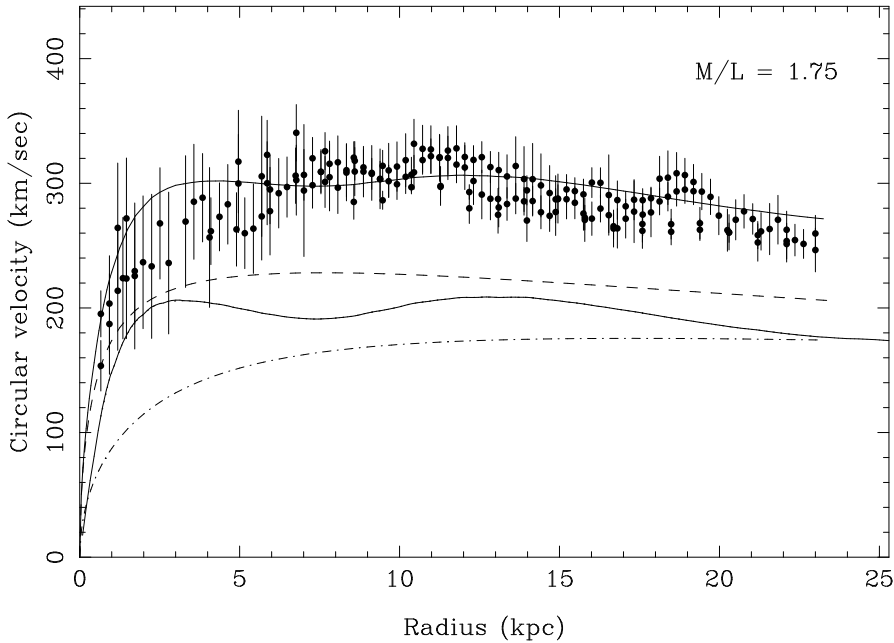}
\caption{\footnotesize Rotation curve of the compressed NFW halo
model.  The dot-dash line shows the uncompressed halo, the dashed line
shows the halo after compression, while the other two lines show the
disk contribution and the total rotation curve.}
\label{comprfig}
\end{figure}

However, these fitted parameters are for the current halo of NGC~1365,
which must have been compressed as the massive disk formed.  We have
therefore adopted the procedure described by \citet{SM05} to try to
determine the parameters of the original uncompressed halo before the
disk formed.  The value $\Upsilon_I=1.75$ for our best fit model
yields a disk mass $M_d = 1.6 \times 10^{11}\;$M$_\odot$, which must
have condensed from the original uncompressed halo to its present
radial distribution.  Assuming the dark matter in the original halo to
have an isotropic velocity distribution, we determine the final mass
profile of the dark matter from some assumed initial NFW form.  We
proceed by trial and error to determine the initial halo that yields a
final rotation curve that closely resembles the fit shown in
right-bottom panel of Fig.~\ref{rotcurvefig} for $\Upsilon_I=1.75$

We find that an initial NFW halo with the properties $c=22$, $V_{200}
= 123\;$\kms, $M_{200}= 5.9 \times 10^{11}\;$M$_\odot$, and
$r_s=8\;$kpc resembles the mass model of the final best-fit NFW halo,
as shown in Figure~\ref{comprfig}.  The inner halo is an excellent
match, but the density of the compressed halo is fractionally higher
than in Fig.~\ref{rotcurvefig} beyond $r=10\;$kpc, which we tolerate
because it errs on the side of allowing a more massive halo.
Nevertheless, the concentration is still high, though less extreme
than if compression is neglected, and $V_{200}$ is still low for such
a massive galaxy.  Furthermore the disk mass fraction, $M_d/M_{200}
\simeq 0.27$ is much greater than the expected value of $\sim 0.05$
\citep[e.g.][]{dutton07}.  Since our estimated value even exceeds the
cosmic dark matter fraction \citep{spergel07}, it may indicate an
unusual formation history for this galaxy.

\subsection{Population synthesis models}
In this study, we have obtained a dynamical estimate of $\Upsilon_I$
for NGC 1365.  In previous work, we obtained estimates by the same
procedure for NGC~4123 \citep{wein2} and NGC~3095 \citep{Ben3095}.
Here we compare our estimated values of $\Upsilon_I$ with the
predictions from population synthesis models by \citet{Bell03} for a
galaxy of the observed B-V color.  Since \citet{Bell03} do not give
predictions for our measured V-I color index, we also compare with the
earlier predictions of \citet{BdJ}.

Table 3 presents the comparison for all three galaxies.  We give
isophotal magnitudes corrected for internal
\citep{giovanelli94,sakai00} and foreground extinction
\citep{schlegel}, and determine the predicted $\Upsilon_I$ from the
B-V color using table 7 of \citet{Bell03}, and from the V-I color
using table 1 of \citet{BdJ}.  \citet{Bell03} suggest the uncertainty
in their prediction is 0.1 dex, or some 25\%.  The final column gives
our dynamically-estimated value of $\Upsilon_I$ for each of the three
galaxies.

Our dynamical estimates of $\Upsilon_I \simeq 2-2.25$ agree reasonably
well with the values of 1.5 -- 2.1 deduced from the stellar population
V-I colors for a ``diet Salpeter'' IMF \citet{BdJ}.  However, our
dynamical modeling estimates are higher than those inferred from B-V
colors \citet{Bell03}, suggesting that these galaxies are somewhat
bluer in B-V for a given V-I than are the stellar population models;
B-V is more sensitive than V-I to the details of recent star
formation.

Our fluid dynamical $\Upsilon_I$ values provide a zeropoint
normalization for the stellar population models that is at or slightly
higher than the preferred normalization of \citet{dJB07}.  The present
work makes this conclusion firmer by removing some distance
uncertainty in $\Upsilon$, since NGC~1365 has a Cepheid distance while
NGC~3095 and NGC~4123 do not.  The $\Upsilon_I$ values from dynamics
make it difficult to have a very low disk, $\Upsilon_I \la 1.0$, as
advocated by some authors to reconcile the Tully-Fisher relation
zeropoint or lack of radius dependence with theoretical expectations
\citep{Portinari,Pizagno}.

\section{Conclusions}
We have presented a detailed study of the strongly-barred galaxy
NGC~1365, including new photometric images and Fabry-Perot
spectroscopy, as well as a detailed re-analysis of the neutral
hydrogen observations by J\"ors\"ater \& van Moorsel (1995).  We find
the galaxy to be at once both remarkably bi-symmetric in its I-band
light distribution and strongly asymmetric in the distribution of dust
and gas, and in the kinematics of the gas.  These asymmetries extend
throughout the galaxy, affecting the bar region, the distribution of
gas in the spiral arms and the neutral hydrogen beyond the edge of the
bright disk.  The velocity field mapped in the \ha\ line showed
bright HII regions with velocities that differed by up to $\sim
80\;$\kms\ from that of the surrounding gas.  Our sparsely-sampled
line profiles in these anomalous velocity regions hint at unresolved
substructure, suggesting a possible double line profile.

The strong bar and spiral arms complicate the determination of the
projection geometry of the disk, assuming it can be characterized as
flat in the inner parts.  The inclination of the plane we derive from
the kinematic data is smaller by about 10$^\circ$ from that determined
from the photometry.  The strong spiral arms that cross the projected
major axis far out in the disk seem likely to bias the photometric
inclination and we therefore adopt, in common with other workers, the
inclination derived from the gas kinematics.  This preference is
supported by the much poorer fits to the observed kinematics obtained
when we adopt the photometric inclination (\S~\ref{comparsect}).

Our attempts to derive the rotation curve of NGC~1365 were complicated
by the fact that neither the \ha\ nor the HI velocity maps are
consistent with a simple circular flow pattern over a significant
radial range.  The bar and spirals clearly distort the gas flow in the
luminous disk.  The neutral hydrogen extends somewhat beyond the
visible disk but unfortunately has neither a uniform distribution nor
regular kinematics.  JvM95 attempted to fit a warp to the outer HI
layer that extends into the visible disk, and derived a strongly
declining rotation curve.  We chose instead to assume a coplanar flow
out to a deprojected radius of 255\arcsec\ and to neglect the
asymmetric velocities in the neutral hydrogen beyond.  The velocities
derived separately from the \ha\ and HI data are in good
agreement.  Our resulting rotation curve shows a gentle decrease
beyond a radius of $\sim 10\;$kpc, similar to those observed in other
massive galaxies \citep{CvG91, noordermeer07}.

We used our deprojected I-band image to estimate the gravitational
field of the luminous matter, which can be scaled by a single
mass-to-light ratio, $\Upsilon_I$.  We also employed a gradual
increase to $\Upsilon_I$ in the central few kpc to allow for an older
bulge-like stellar population, although the light distribution does
not appear to have a substantially greater thickness near the center.
We combined the central attraction of the axially-symmetrized disk for
various values of $\Upsilon_I$ with two different halo models to fit
the observed rotation curve in the region outside the bar -- finding,
as always, no significant preference for any $\Upsilon_I$.

We attempt to fit hydrodynamic simulations of the gas flow pattern in
the bar region, in order to constrain $\Upsilon_I$.  For each type of
halo adopted, we run a grid of simulations covering a range of both
$\Upsilon_I$ and $\Omega_p$, the pattern speed of the bar.  We then
project each simulation to our adopted orientation of the galaxy and
compare the gas flow velocities in the model with those observed.
Since the light distribution in NGC~1365 is highly symmetric, our
simulations were constrained to be bi-symmetric, yet the observed gas
flow has strong asymmetries.  None of our simulations is capable,
therefore, of fitting both sides of the bar simultaneously.

The anomalous position of the dust lane in the western part of the bar
suggests that side is the more likely to be disturbed, and we
therefore fit our models to the eastern half of the bar only.  After
smoothing the model to match the resolution of the kinematic data and
masking out five blobs of gas with strongly anomalous velocities, we
are able to obtain moderately satisfactory fits to the remaining
velocities.  The best fit pattern speed is $\Omega_p =
24\;$\kms$\rm{kpc}^{-1}$ for both types of halo which places
corotation for the bar at $r_L \simeq 1.23r_B$, in excellent agreement
with the value found by \citet{Linlinatha} and consistent with most
determinations of this ratio for other galaxies \citep[\eg][]{ADC03}.

Our estimated mass-to-light ratio values are $\Upsilon_I \simeq 2.50
\pm 1$ for the isothermal halo models and $\Upsilon_I=1.75 \pm 1$ for
the NFW halo.  While the constraints are disappointingly loose, the
preferred mass-to-light ratio in both our halo models, $\Upsilon_I
\simeq 2.0 \pm 1$, is consistent with that obtained by \citet{wein2}
and \citet{Ben3095} in two other cases.  For NGC~1365, however, this
value implies a massive, but not fully maximal, disk, and we do not
find support for the disk-only model with no halo that was suggested
by JvM95.  Although such a model can reproduce the declining rotation
curve (see their Fig. 24), the simulated gas flow produced by such a
model (which in the I band is $\Upsilon\sim 3.75$) is quite strongly
excluded.

The preferred value of $\Upsilon_I$ is nicely consistent with those
obtained in the two previous studies using this method \citep{wein2,
Ben3095}, but suggest somewhat more massive disks than predicted by
population synthesis models \citep{BdJ,Bell03} for galaxies of these
colors.

The halos of our two models required to fit the declining rotation
curve in the outer disk are distinctly non-standard, however.  The
circular speed in the pseudo-isothermal model declines steadily
outside the large core, while the NFW halo has a very high
concentration and small scale radius.  Even allowing for compression
of the halo as the massive disk forms within it, the original NFW halo
has $c \simeq 22$, $V_{200} \simeq 123\;$kms.  The total dark matter
mass out to $r_{200}$ in this case is less than three times our
estimated disk mass, and the halo is quite unlike those predicted by
LCDM models for a galaxy of this mass.

The disturbed distribution and kinematics of the gas in this galaxy
clearly complicates our attempt to identify a preferred mass model.
Its projected position near the center of the Fornax cluster, together
with its velocity within 200~\kms\ of the cluster mean, suggest it is
a cluster member.  The disturbed nature of the outer HI distribution
should not therefore be regarded as surprising.  But the high central
density of this massive galaxy should ensure that tidal forces have
little influence on the inner part, where we indeed see that the bar
and inner spirals are very pleasingly bi-symmetric in the I-band
light.  The existence of such strong asymmetries in the inner parts of
the gas and dust is rather surprising, therefore.

The asymmetry in the dust distribution and the kinematic map, combined
with the existence of a number of patches of \ha\ emission with
anomalous velocities all suggest that the agent that caused the
disturbance was an infalling gas cloud.  We cannot say whether the gas
was an isolated intergalactic cloud not associated with a galaxy, or
whether it could be a stream of debris from a gas-rich dwarf galaxy
that had been tidally disrupted.  The anomalous velocities clearly
suggest that the infalling gas has yet to be assimilated in the disk of
NGC~1365.

\acknowledgments
We are grateful to Scott Trager for obtaining photometric images at
the Las Campanas Observatory.  We received invaluable help and advice
with the optical data reductions from Tad Pryor and with the radio
data reductions from Gustaaf van Moorsel and the NRAO staff.  This
work was supported by NSF grants AST-0098282 and AST-0507323 awarded
to JAS.


\clearpage

\begin{deluxetable}{crrrr}
\tabletypesize{\scriptsize}

\tablecaption{Halo Parameters \label{halotab}}
\tablecolumns{5}
\tablewidth{0pt}
\tablehead{
\colhead{Disk $\Upsilon_I$} & \colhead{$\rho_0$} & \colhead{$r_c$} &  \colhead{$k$} &
\colhead{$\chi^2/N$}
}
\startdata
\colhead{} & \multicolumn{4}{c}{b1025} \\
\cline{2-5} \\
1.00 & 1433.20 & 1.45 & 2.61& 0.92 \\
1.50 &  420.20 & 2.64 & 3.00 & 0.92 \\
2.00 & 184.70 & 3.72 & 3.50 & 0.91 \\
2.50 & 81.20 &  5.16 & 4.78 & 0.93 \\
3.00 & 67.20 & 4.60 & 5.0 & 0.99  \\

\colhead{} & \multicolumn{4}{c}{b1075} \\
\cline{2-5} \\
0.50 & 1422.16 & 1.58 & 2.62 & 0.97\\
0.75 & 784.56 & 2.11 & 2.77 & 0.96 \\
1.00 & 346.08 & 3.19 & 3.10 & 0.92 \\
1.25 & 354.20 & 3.01 & 3.05 & 0.87 \\
1.50 & 241.21 & 3.55 & 3.27 & 0.86 \\
1.75 & 173.99 & 4.02 & 3.51 & 0.86 \\
2.00 & 126.70 & 4.51 & 3.81 & 0.87 \\
2.25 & 91.30 & 5.06 & 4.29 & 0.89 \\
2.50 & 62.55 & 5.72 & 5.0 & 0.92 \\
2.75 & 56.43 & 5.41 & 5.0 & 0.97 \\
3.00 & 49.10 & 4.87 & 5.0 & 1.06 \\
3.25 & 49.92 & 4.06 &  5.0 & 1.21 \\
3.50 & 18.16 & 3.87 &  5.0 & 1.36 \\
3.75 & 2.02  & 1.36 &  5.0 & 1.70 \\

\colhead{} & \multicolumn{4}{c}{b2050} \\
\cline{2-5} \\
1.00 & 691.78 & 2.13 & 2.76 & 0.92 \\
1.50 & 212.37 & 3.75 & 3.31 & 0.89 \\
2.00 & 109.76 & 4.75 & 3.86 & 0.89 \\
2.50 & 53.03 & 5.97 & 5.0 & 0.95 \\
3.00 & 36.85 & 5.18 & 5.0 & 1.08 \\

\colhead{} & \multicolumn{4}{c}{b2075} \\ 
\cline{2-5} \\
1.00 & 381.17 & 2.96 & 2.98 & 0.90 \\
1.50 & 166.99 & 4.20 & 3.44 & 0.86 \\
2.00 & 90.10 & 5.14 & 3.98 & 0.90 \\
2.50 & 43.60 & 6.31 & 5.0 & 1.00 \\
3.00 & 30.49 & 5.23 & 5.0& 1.18 \\

\colhead{} & \multicolumn{4}{c}{b3075} \\
\cline{2-5} \\
1.00 & 305.24 & 3.29 & 3.05 & 0.92 \\
1.50 & 134.05 & 4.59 & 3.52 & 0.88 \\
2.00 & 82.60 & 5.06 & 3.76 & 0.92 \\
2.50 & 32.46 & 6.63  & 5.0 & 1.05 \\
3.00 & 14.77 & 5.85 & 5.0& 1.26 \\
\cline{2-5}\\
\enddata

\tablecomments{Dark matter halo parameters for five different bulge
models.  For models with $\Upsilon_I = 2.75$ and
3.00, the k parameter was fixed to 5 to avoid a sharp edges on the 
halo profile}

\end{deluxetable}

\begin{deluxetable}{crrrrr}
\tabletypesize{\scriptsize}

\tablecaption{NFW Halo Parameters \label{halotabNFW}}
\tablecolumns{6}
\tablewidth{0pt}
\tablehead{
\colhead{Disk $\Upsilon_I$} & \colhead{$\rho_s$} & \colhead{$r_s$} &
\colhead{$c$} & \colhead{$V_{200}$} & \colhead{$\chi^2/N$}
}
\startdata
0.50 & 829.96 & 2.90 & 67 & 135 & 0.92 \\
0.75 & 802.80 & 2.84 & 66 & 131 & 0.93 \\
1.00 & 710.48 & 2.87 & 63 & 127 & 0.90 \\
1.25 & 685.32 & 2.80 & 62 & 122 & 0.92 \\
1.50 & 662.08 & 2.71 & 61 & 116 & 0.93 \\
1.75 & 641.40 & 2.61 & 61 & 111 & 0.96 \\
2.00 & 563.16 & 2.61 & 58 & 106 & 0.98 \\
2.25 & 182.60 & 3.94 & 38 & 105 & 1.16 \\
2.50 & 190.40 & 3.55 & 38 &  96 & 1.15 \\
2.75 & 208.36 & 3.08 & 40 &  86 & 1.15 \\
3.00 & 240.20 & 2.54 & 42 &  75 & 1.17 \\
3.25 & 229.16 & 2.22 & 41 &  64 & 1.26 \\
3.50 & 173.00 & 1.85 & 37 &  48 & 1.33 \\
3.75 &   0.36 & 2.17 &  3 &   4 & 1.42 \\
\enddata

\tablecomments{Dark matter halo parameters for NFW halo. The densities are in
$10^{-3}M_{\sun}pc^{-3}$, the radii in kpc, and $V_{200}$ is in \kms. }

\end{deluxetable}

\begin{deluxetable}{crrrccc}
\tabletypesize{\scriptsize}

\tablecaption{Comparison with Population Synthesis}
\tablecolumns{7}
\tablewidth{0pt}
\tablehead{
\colhead{Galaxy} & \colhead{B} &  \colhead{V} &  \colhead{I} &
\colhead{Pred 1 $\Upsilon_I$} & \colhead{Pred 2 $\Upsilon_I$} &
\colhead{Dynamical $\Upsilon_I$}
}
\startdata
NGC 4123 & 11.97 & 11.38 & 10.36 & 1.22 & 1.48 & $2.25\pm0.25$ \\
NGC 3095 & 11.72 & 11.16 & 10.10 & 1.16 & 1.67 & $2.00\pm0.25$ \\
NGC 1365 &       &  9.21 &  8.05 &      & 2.28 & $2.00\pm1.00$ \\
\enddata

\tablecomments{Our galaxy isophotal magnitudes in 3 color bands corrected
for internal and foreground extinction.  Prediction 1 $\Upsilon_I$ is from
B-V color using \citet{Bell03} table 7.  Prediction 2 $\Upsilon_I$ from
V-I color using \citet{BdJ} table 1. Our dynamical estimate of
$\Upsilon_I$. }

\end{deluxetable}

\end{document}